\documentclass[a4paper,11pt]{article}
\usepackage{jheppub} 
\usepackage{mathtools}
\usepackage[capitalize]{cleveref}
\usepackage[modulo]{lineno}
\usepackage{siunitx}
\usepackage{xcolor}
\usepackage{hhline}
\usepackage{enumitem}

\DeclarePairedDelimiter\ket{\lvert}{\rangle}
\DeclarePairedDelimiterX\braket[2]{\langle}{\rangle}{#1\,\delimsize\vert\,\mathopen{}#2}

\title{\boldmath Constraints on non-unitary neutrino mixing in light of atmospheric and reactor neutrino data}

\author[a]{Tetiana Kozynets}
\author[b]{Philipp Eller}
\author[b]{Alan Zander}
\author[c,d]{Manuel Ettengruber}
\author[a]{D. Jason Koskinen}
\affiliation[a]{Niels Bohr Institute -- University of Copenhagen,\\Jagtvej 155A, 2200 Copenhagen, Denmark}
\affiliation[b]{Technical University of Munich (TUM),\\James-Franck-Strasse 1, 85748 Garching, Germany}
\affiliation[c]{
	Arnold Sommerfeld Center,
	Ludwig-Maximilians-Universit{\"a}t,
	Theresienstra{\ss}e 37,
	80333 M{\"u}nchen,
	Germany,}
\affiliation[d]{Max-Planck-Institut f\"{u}r Physik (Werner-Heisenberg-Institut), F\"{o}hringer Ring 6,\\80805 Munich, Germany}

\emailAdd{tetiana.kozynets@nbi.ku.dk}

\abstract{While the origin of neutrino masses remains unknown, several key neutrino mass generation models result in a non-unitary three-neutrino mixing matrix. To put such models to test, the deviations of the mixing matrix from unitarity can be measured directly through neutrino oscillation experiments. In this study, we perform a Bayesian analysis of the non-unitary mixing model using the recent public data from atmospheric and reactor neutrino experiments — namely IceCube-DeepCore, Daya Bay, and KamLAND. The novelty of our approach compared to the preceding global fits for non-unitarity is in the detailed treatment of the atmospheric neutrino data, which for the first time includes the relevant flux and detector systematic uncertainties. From the Bayesian posteriors on the individual mixing matrix elements, we derive the non-unitarity constraints in the form of normalisations and closures of the mixing matrix rows and columns, assuming either a fully unconstrained matrix or a physically motivated submatrix scenario. We find comparable constraints for electron and tau row normalisations as other similar studies in literature, and additionally reveal strong correlations between muon and tau row constraints induced by the atmospheric systematic uncertainties. We find that the current data is well described by both unitary and non-unitary mixing models, with a strong preference for the unitary mixing indicated by the Bayes factor. With the upcoming IceCube-Upgrade and JUNO detectors, both featuring superior energy resolution compared to the current atmospheric and reactor neutrino experiments, our constraints on the row normalisations in the submatrix case are expected to improve by 25\%, 40\%, and 20\% in the electron, muon, and tau sectors respectively. In the future, our approach can be expanded to include solar and long-baseline neutrino experiments, with the aim to provide more stringent constraints while keeping track of the nuisance parameters that may be degenerate with non-unitarity. }

\begin{document}
\maketitle
\flushbottom


\section{Introduction}
\label{sec:intro}

The existence of three generations of the weak force doublets, each comprising a charged lepton $l_{\alpha}$ and a neutrino of the corresponding flavour $\nu_{\alpha}$ ($\alpha \in \{e, \mu, \tau\}$), is one of the fundamental components of the Standard Model (SM). The experimentally observed phenomenon of neutrino flavour transitions ($\nu_{\alpha} \to \nu_{\beta}$) implies mixing between three propagating neutrino states $\nu_i$ ($i \in \{1, 2, 3\}$) of non-degenerate, and, consequently, non-zero masses $m_i$\footnote{At least two of the three mass states are required to have non-zero masses.}. At present, the latter are not described as part of the SM, and the origin of neutrino masses remains unknown. Several neutrino mass models (for a review, see \cite{deGouvea:2016qpx,Ma:2009dk}) suggest the existence of a new heavy mass scale $M$, $M \lesssim \Lambda = \mathcal{O}(10^{14}\,\mathrm{GeV})$, whose effects in a low-energy ($\ll M$) theory can be captured via addition of higher-dimensional ($\dim > 4$) operators to the SM Lagrangian. The lowest-dimension operator of such kind, the dim-5 Weinberg operator $\mathcal{L}_5$, gives rise to parametrically suppressed neutrino masses $m_{\nu} \propto \frac{v^2}{\Lambda}$, where $v \simeq 246\,\mathrm{GeV}$ is the vacuum expectation value of the SM Higgs (``electroweak scale''). This parametric suppression explains the relative smallness of the neutrino masses compared to those of other fermions, making the Weinberg operator an appealing effective description of neutrino mass generation at energies much lower than the new mass scale $M$ \cite{deGouvea:2016qpx}. In certain variants of a complete high-energy theory, $M$ can be interpreted as the Majorana mass of additional right-handed neutrino states, such as in the type-I (type-III) seesaw mechanism, which features the addition of an SM gauge singlet (triplet) to the Standard Model. The heavy right-handed neutrino models are also hypothesised in the context of neutrino origin of dark matter, which could consist, fully or in part, of such “sterile” neutrinos (see e.g. \cite{Dev:2016qbd,Dev:2016qeb,Boyarsky:2018tvu}).

The particle physics and astrophysics motivation for the existence of extra neutrino states raises a broader question of unitarity of the active three-neutrino mixing matrix (also known as the Pontecorvo–Maki–Nakagawa–Sakata, or PMNS, matrix). In the unitary scenario postulated for the PMNS matrix, the total probability for one of the $\{e, \mu, \tau\}$ flavour states to oscillate to either one of these three flavours is conserved and equal to 1. However, this need not be the case in the presence of mixing of the light active states with the heavy right-handed ones, which would make the $3 \times 3$ PMNS matrix a non-unitary submatrix of a larger unitary matrix. As a consequence, the $\ell^2$-norms (``normalisations'') of the rows and the columns of the $3 \times 3$ matrix would be reduced to values smaller than 1, affecting neutrino oscillations, production, and detection \cite{Antusch:2006vwa,Blennow:2016jkn,Ellis:2020hus}. 

In this study, we focus on the regime of ``minimal unitarity violation'' (MUV) \cite{Antusch:2006vwa,Ellis:2020hus}, where non-unitarity is induced by physics that exists at energy scales much larger than $v$. In this regime, percent and sub-percent constraints on leptonic non-unitarity have been placed via e.g. searches for flavour-violating decays of charged leptons, electroweak universality tests, measurements of the invisible $Z$-boson decay width, and other searches detailed in \cite{deGouvea:2015euy,Antusch:2014woa,Fernandez-Martinez:2016lgt,Blennow:2023mqx}. At the same time, neutrino oscillation data provides a way to study the unitarity of neutrino mixing directly, without invoking complementary channels or constraints from the electroweak sector. This approach has been taken in e.g. \cite{Parke:2015goa,Ellis:2020hus,Denton:2021mso,Forero:2021azc}, where neutrino oscillation amplitudes were inferred from published experimental results and reinterpreted as constraints on the individual mixing matrix elements and combinations thereof. Such constraints were typically derived as the result of global fit studies, in which neutrino experiments with different baselines, energy ranges, and oscillation channels were combined to probe the entire neutrino mixing matrix. All of short- and long-baseline, solar, reactor, and atmospheric neutrino experiments were fully or partially incorporated in the literature dedicated to global fits \cite{Parke:2015goa,Ellis:2020hus,Ellis:2020ehi,Forero:2021azc}, and projections for selected next-generation experiments have similarly been made \cite{Agarwalla:2021owd,Escrihuela:2016ube,Sahoo:2023mpj}.

The main goal of the present study is to draw attention to the treatment of atmospheric neutrino data in the non-unitarity analyses, which, to the best of our knowledge, has so far been incomplete. Although several preceding studies \cite{Parke:2015goa,Denton:2021mso} reinterpreted the Super-Kamiokande results \cite{Super-Kamiokande:2010orq,Super-Kamiokande:2014ndf,Super-Kamiokande:2017edb} and the IceCube-DeepCore oscillation results \cite{IceCube:2017lak,IceCube:2019dqi} to constrain unitarity in the muon and the tau rows of the mixing matrix, these analyses have not considered atmospheric neutrino systematic uncertainties as prescribed by the respective collaborations. As we show in this paper, the latter are crucial for placing accurate non-unitarity constraints, since the nuisance parameters may introduce energy- and direction-dependent effects correlated with those of the non-unitarity physics. In particular, the relevant uncertainties include those related to the unoscillated atmospheric neutrino flux, such as the overall normalization, spectral index, $\nu/\bar{\nu}$ and $\nu_e$/$\nu_{\mu}$ ratios, and angular distributions\footnote{See \cite{Barr:2006it} for a simplified treatment and \cite{Yanez:2023lsy,Fedynitch:2022vty} for a refined derivation of the atmospheric neutrino flux systematic uncertainties.}, as well as detector-specific systematics. We seek to address these uncertainties in the context of non-unitarity constraints for the first time and study their degeneracies with the non-unitarity metrics (such as the normalisations of the mixing matrix rows and columns). We stress that fitting for the non-unitarity physics parameters and the experimental systematic parameters at the same time is a significantly more robust approach than the reinterpretation of the best-fit constraints from the three-flavour analyses, since the best-fit systematic parameters might already be absorbing some non-unitarity effects.

The approach we propose above is possible with the  public atmospheric neutrino data and Monte Carlo simulation from the IceCube-DeepCore experiment \cite{IceCube:2019pdr}, which has not been considered in literature as part of any global fit studies focused on non-unitarity\footnote{Although this dataset corresponds to the previous generation of the IceCube-DeepCore analyses \cite{IceCube:2017lak,IceCube:2019dqi} with only $\sim$3 years of data, and improved analyses with $\sim$3 times more data and a refined treatment of systematic uncertainties have since been published \cite{IceCube:2023ins,IceCubeCollaboration:2023wtb}, it remains the only publicly available atmospheric neutrino dataset that contains all of the information necessary for performing a non-unitarity analysis. The recent public data release from the Super-Kamiokande experiment \cite{SuperKamiokande:2023pdr} does not provide any prescriptions for implementing the systematic uncertainties, which is why we are not considering it in this study.}. This dataset covers atmospheric muon neutrino disappearance and tau neutrino appearance channels, thereby providing access to the elements of the muon and the tau rows of the neutrino mixing matrix. To form a ``minimal''  selection of datasets that would let us probe all three matrix rows, we supplement the IceCube-DeepCore data with reactor neutrino data from Daya Bay \cite{DayaBay:2022orm} and KamLAND \cite{KamLAND:2008dgz} experiments. The reactor experiments provide a handle on the elements of the electron row of the mixing matrix, whose measurements are similarly subject to the reactor systematic uncertainties and are implemented in this study to the extent possible with the publicly available information. To assess how the non-unitarity constraints possible with this minimal selection of experiments will evolve in the future, we further develop an equivalent analysis including the next-generation IceCube-Upgrade \cite{Ishihara:2019aao} and JUNO \cite{JUNO:2022mxj} experiments (both in deployment at the time of writing). For these future projections, we once again perform the global fit for the individual mixing matrix elements not constrained by unitarity alongside the systematic parameters known at this stage. While the future constraints will evolve as the experiments become operational and settle on the event selection and data analysis pipelines, our study provides 
the first attempt to utilize the preliminary IceCube-Upgrade simulation in this context \cite{IceCube:2020umc} and serves as a proof of concept for the upcoming studies with an improved detector simulation and the upcoming data.

This paper is structured as follows. In \cref{sec:formalism}, we review the non-unitary neutrino mixing formalism, covering the impact of non-unitarity on neutrino oscillation probabilities as well as the flux and cross section normalisation effects. \cref{sec:experiments} provides details on the experimental datasets included in this study and shows the impact of the deviations from unitarity on the expected event templates. In \cref{sec:validation}, we validate our analysis setup for each experiment by reproducing the standard three-flavour oscillation results from the respective collaborations. \cref{sec:global_fit_setup} proceeds with the description of the Bayesian global fit we employ to derive the current constraints on the individual mixing matrix elements and the ensuing non-unitarity metrics. Our main results are given in \cref{sec:main_results_current}. In \cref{sec:atmo_sys_impact}, we further show how the atmospheric neutrino systematic uncertainties influence the derived posterior distributions for the non-unitarity metrics. Future projections for the next-generation atmospheric and reactor neutrino experiments are given in \cref{sec:future_projections}. Finally, we discuss the implications of our findings and make suggestions for further advancements of this analysis in \cref{sec:discussion}.

\section{Non-unitary neutrino mixing formalism}\label{sec:formalism}
\subsection{Mixing matrix parameterisation}
In the standard three-flavour neutrino oscillation framework, flavour states $\ket{\nu_{\alpha}}$ produced together with the respective charged leptons are related to the propagating mass states $\ket{\nu_i}$ via a unitary PMNS matrix $U$:

\begin{equation}
    \ket{\nu_{\alpha}} = \sum_{i=1}^{3} U^{*}_{\alpha i} \ket{\nu_{i}},
\end{equation}
where $\alpha \in \{e, \mu, \tau\}$. The PMNS matrix is usually parameterised as a product of three rotation matrices \cite{Workman:2022ynf}:

\begin{equation}
\begin{split}
    U &=  \begin{pmatrix}
    1 & 0 & 0 \\
    0 & c_{23} & s_{23} \\
    0 & -s_{23} & c_{23}
    \end{pmatrix} \begin{pmatrix}
    c_{13} & 0 & s_{13}e^{-i\delta_{\mathrm{CP}}} \\
    0 & 1 & 0 \\
    -s_{13}e^{i\delta_{\mathrm{CP}}} & 0 & c_{13}
    \end{pmatrix} \begin{pmatrix}
    c_{12} & s_{12} & 0 \\
    -s_{12} & c_{12} & 0 \\
    0 & 0 & 1
    \end{pmatrix},
\end{split}
\label{eq:unitary_matrix_mixing_angles}
\end{equation}
where $\delta_{\mathrm{CP}}$ is the physical Dirac CP-violating phase, $c_{ij} \equiv \cos \theta_{ij}$, $s_{ij} \equiv \sin \theta_{ij}$, and $\theta_{ij}$ is the mixing angle between mass eigenstates $i$ and $j$. As discussed in \cref{sec:intro}, the existence of new physics (such as extra neutrino states) may render the $3 \times 3$ neutrino mixing matrix non-unitary. In the most general form, such a matrix may be parameterised as
\begin{equation}
    N = \begin{pmatrix} |N_{e1}| & |N_{e2}|e^{i\phi_{e2}} & |N_{e3}|e^{i\phi_{e3}} \\ |N_{\mu1}| & |N_{\mu2}|\phantom{e^{i\phi_{\mu x}}} & |N_{\mu3}|\phantom{e^{i\phi_{\mu x}}} \\ |N_{\tau1}| & |N_{\tau2}|e^{i\phi_{\tau2}} & |N_{\tau3}|e^{i\phi_{\tau3}}\end{pmatrix},    
    \label{eq:generic_mixing_matrix}
\end{equation}
where both $NN^{\dag}$ and $N^{\dag}N$ may deviate from identity. The parameterisation \eqref{eq:generic_mixing_matrix} includes 9 real non-negative matrix element magnitudes $|N_{\alpha i}|$ and 4 complex phases $\phi_{\alpha i}$, which can be assigned to any $2 \times 2$ submatrix of $N$ \cite{Ellis:2020hus}. 

In the MUV case considered in this study, only the usual low-mass neutrino mass eigenstates $\ket{\nu_i}$ with masses $m_i$ ($i \in {1, 2, 3}$) are kinematically accessible in an experiment. The effective neutrino flavour state $\ket{\nu_{\alpha}^{\mathrm{eff.}}}$ at the time of production or detection can only be a superposition of these accessible mass states, i.e.,

\begin{equation}
    \ket{\nu_{\alpha}^{\mathrm{eff.}}} = \frac{1}{\sqrt{(NN^{\dag})_{\alpha\alpha}}} \sum_{i=1}^{3}N^{*}_{\alpha i} \ket{\nu_{i}} \equiv \frac{1}{\sqrt{N_{\alpha}}} \sum_{i=1}^{3}N^{*}_{\alpha i} \ket{\nu_{i}},
\label{eq:effective_flavour_states}
\end{equation}
where the sum must be truncated at $i = 3$. The normalisation factor $(\sqrt{(NN^{\dag})_{\alpha \alpha}})^{-1}$ ensures that $\braket{\nu_{\alpha}^{\mathrm{eff.}}}{\nu_{\alpha}^{\mathrm{eff.}}} = 1$ \cite{Antusch:2006vwa,Aloni:2022ebm,Ellis:2020hus}. However, under a non-unitary mixing matrix $N$, the set of effective flavour states is not orthonormal, i.e., $\braket{\nu_{\beta}^{\mathrm{eff.}}}{\nu_{\alpha}^{\mathrm{eff.}}} \neq \delta_{\alpha \beta}$. An orthonormal basis could be defined in a complete high-energy theory covering the new physics energy scale \cite{Antusch:2006vwa}.

\subsection{Neutrino oscillations in matter}\label{sec:nonunitary_osc_in_matter}

The propagation of the three mass states $\ket{\nu_i}$ in matter is governed by the following Hamiltonian (expressed in the mass basis):
\begin{equation}
    H = H_{\mathrm{vacuum}} + A = \frac{1}{2E}\begin{pmatrix}
    0 & 0 & 0 \\ 
    0 & \Delta m^2_{21} & 0 \\
    0 & 0 & \Delta m^2_{31}\end{pmatrix} + 
    N^{\dag}\begin{pmatrix}
        V_{\mathrm{CC}} + V_{\mathrm{NC}} & 0 & 0 \\
        0 & V_{\mathrm{NC}} & 0 \\
        0 & 0 & V_{\mathrm{NC}}
    \end{pmatrix} N.
\label{eq:matter_hamiltonian_mass_basis}
\end{equation}
The first term in \cref{eq:matter_hamiltonian_mass_basis} corresponds to the free propagation in vacuum, where $\Delta m_{ij}^2 \equiv m^2_j - m^2_i$ are the mass splittings between the physical states $i$ and $j$. The second term is the contribution of the matter potential $A$ due to the charged current (CC) and the neutral current (NC) interactions of neutrinos with electrons and nucleons in matter. As electrons are the only leptons that compose the ordinary stable matter, the CC interactions via $W^{\pm}$ exchange are accessible only to electron neutrinos and antineutrinos. The $V_{\mathrm{CC}}$ component of the matter potential thus only depends on electron number density $n_e$:
\begin{equation}
    V_{\mathrm{CC}} = \pm \sqrt{2} G_{\mathrm{F}} n_e = \pm \sqrt{2} G_{\mathrm{F}} E_{\nu} \rho_m Y_e N_A,
\label{eq:charged_current_potential}
\end{equation}
where $G_{\mathrm{F}}$ is the Fermi constant, $\rho_m$ is the mass density of the medium in $\mathrm{g}\,\mathrm{cm}^{-3}$, $Y_e$ is the electron fraction per unit molar mass (also known as the ``$Z/A$'' factor and expressed in $\mathrm{mol}\,\mathrm{g}^{-1}$), and $N_A$ is the Avogadro number. In \cref{eq:charged_current_potential}, the ``+'' sign corresponds to the matter potential experienced by neutrinos ($\nu_{\alpha}$), and the ``-'' sign -- that seen by antineutrinos ($\bar{\nu}_{\alpha}$). The neutral current ($Z^{0}$-exchange) interactions of neutrinos could occur with protons, neutrons, and electrons alike. However, a typical assumption is that the medium that neutrinos travel through is neutral and unpolarized, in which case the contributions due to proton and electron NC potentials cancel each other out \cite{Wallraff:2014qka,Escrihuela:2016ube,Blennow:2013rca}. The only remaining component is the flavour-independent neutrino-neutron scattering, with the respective NC potential:
\begin{equation}
    V_{\mathrm{NC}} = \mp \frac{\sqrt{2}}{2} G_{\mathrm{F}} n_n = \mp \frac{\sqrt{2}}{2} G_{\mathrm{F}} \rho_m Y_n N_A,
\label{eq:neutral_current_potential}
\end{equation}
where $Y_n = 1 - Y_e$ is the neutron fraction per unit molar mass, and the rest of the notations are the same as in \cref{eq:charged_current_potential}. In the unitary case, the neural current potential is usually omitted in \cref{eq:matter_hamiltonian_mass_basis}, as it is identical for all active neutrino flavours and contributes only an unobservable phase to the neutrino oscillation amplitude \cite{Blennow:2013rca}. This is not the case when the mixing matrix is non-unitary, such that $N^{\dag} N \neq \mathbb{I}$, which necessitates the explicit inclusion of $V_{\mathrm{NC}}$ in \cref{eq:matter_hamiltonian_mass_basis}.

With the Hamiltonian defined as in \cref{eq:matter_hamiltonian_mass_basis}, the Schr\"{o}dinger equation for the propagating neutrino states is:
\begin{equation}
    i\hbar\frac{\partial}{\partial t} \boldsymbol{\nu}_{\mathrm{m}}(t)  = H\boldsymbol{\nu}_m (t),
\label{eq:schroedinger_equation}
\end{equation}
where $\boldsymbol{\nu}_{\mathrm{m}}(t)$ describes the time dependence of the state vector $\boldsymbol{\nu}_m \equiv (\nu_1, \nu_2, \nu_3)^{\top}$. This dependence can be expressed through the evolution operator $S^0$, such that
\begin{equation}
    \boldsymbol{\nu}_{\mathrm{m}}(t) = S^0 (t) \boldsymbol{\nu}_{\mathrm{m}}(0).
\label{eq:time_evolution_operator}
\end{equation}
Plugging \cref{eq:time_evolution_operator} into \cref{eq:schroedinger_equation}, we find 
\begin{equation}
    i\hbar \frac{\partial S^0}{\partial t} = HS^0,
\label{eq:time_derivative_of_evol_operator}
\end{equation}
where $\boldsymbol{\nu}_{\mathrm{m}}(0)$ is eliminated since \cref{eq:schroedinger_equation} holds for any choice of initial condition. When the densities of electrons and neutrons are constant throughout the distance $L$ propagated by the neutrinos in time $t$, we can easily solve for the time dependence of the evolution operator \cite{Barger:1980tf,Blennow:2016jkn,Arguelles:2022tki}: 
\begin{equation}
    S^0(t) \to S^0(L) =  \exp{(-iHL)}.
\end{equation}
The oscillation probabilities from flavour $\alpha$ to flavour $\beta$ can then be found as follows:
\begin{equation}
    P_{\alpha\beta}(E, L) = \frac{|(N S^0(E, L) N^{\dag})_{\beta\alpha}|^2}{(NN^{\dag})_{\alpha\alpha}(NN^{\dag})_{\beta\beta}} = \frac{|(N e^{-iHL} N^{\dag})_{\beta\alpha}|^2}{(NN^{\dag})_{\alpha\alpha}(NN^{\dag})_{\beta\beta}} \equiv \frac{|(N e^{-iHL} N^{\dag})_{\beta\alpha}|^2}{N_{\alpha} N_{\beta}},
\label{eq:nonunitary_osc_probabilities}
\end{equation}
where the $N_{\alpha}$, $N_{\beta}$ factors in the denominator originate from the normalisation of the effective flavour states as per \cref{eq:effective_flavour_states}. 

\subsection{Non-unitary oscillation probabilities}\label{sec:oscillograms_and_normalisation_effects}

The non-unitarity of the $N$ matrix can be quantified as any deviation of $(NN^{\dag})_{ij}$ or $(N^{\dag}N)_{ij}$ from $\delta_{ij}$. Specific cases include off-nominal normalisations of any of the matrix rows ($N_{\alpha} = \sum_{i}|N_{\alpha i}|^2 \neq 1$) or columns ($N_{i} = \sum_{\alpha}|N_{\alpha i}|^2 \neq 1$), as well as non-zero closures of rows ($t_{\alpha\beta} = \sum_{i}N_{\alpha i}^{*} N_{\beta_i} \neq 0$) or columns ($t_{kl} = \sum_{\alpha}N_{\alpha k}^{*} N_{\alpha l} \neq 0$). Such deviations directly affect not only neutrino oscillations as per \cref{eq:nonunitary_osc_probabilities}, but also production and detection of neutrinos \cite{Antusch:2006vwa,Blennow:2016jkn,Ellis:2020hus,Denton:2021mso,Aloni:2022ebm}. This requires treating carefully the flux and the cross section inputs to the projected number of events in an experiment. In particular, if the unoscillated flux $\Phi_{\alpha}$ of the initial neutrino flavour $\alpha$ or the interaction cross section $\sigma_{\beta}$ of the final flavour $\beta$ are based on Standard Model (``SM'') calculations assuming unitarity, then they need to be corrected by appropriate combinations of $N_{\alpha}$ or $N_{\beta}$ in the non-unitary (``NU'') case. In particular,
\begin{subequations}
\begin{align}
    &\Phi_{\alpha}^{\mathrm{NU}} = N_{\alpha} \Phi_{\alpha}^{\mathrm{SM}};\label{eq:normalisation_corr_flux}\\
    &\sigma_{\beta}^{\mathrm{NU,\,CC}} = N_{\beta} \sigma_{\beta}^{\mathrm{SM}};\label{eq:normalisation_corr_cc_xsec}\\
    &\sigma_{\beta}^{\mathrm{NU,\,NC}} = N_{\beta}^2 \sigma_{\beta}^{\mathrm{SM}},\label{eq:normalisation_corr_nc_xsec}
\end{align}
\label{eq:normalisation_corrections}
\end{subequations}
where ``CC'' stands for charged current, and ``NC'' -- for neutral current interactions. We refer the reader to \cite{Antusch:2006vwa,Blennow:2016jkn,Ellis:2020hus,Denton:2021mso,Aloni:2022ebm} for details and derivations of the prefactors in \cref{eq:normalisation_corrections}. We note that these prefactors can be absorbed in the definition of the oscillation probability itself, such that the ``effective'' oscillation probability $\hat{P}_{\alpha\beta}$ is the product of ${P}_{\alpha\beta}$ from \cref{eq:nonunitary_osc_probabilities} and the appropriate factors from \cref{eq:normalisation_corrections}.
 
\begin{figure}[htb!]
    \centering
    \includegraphics[width=0.95\textwidth]{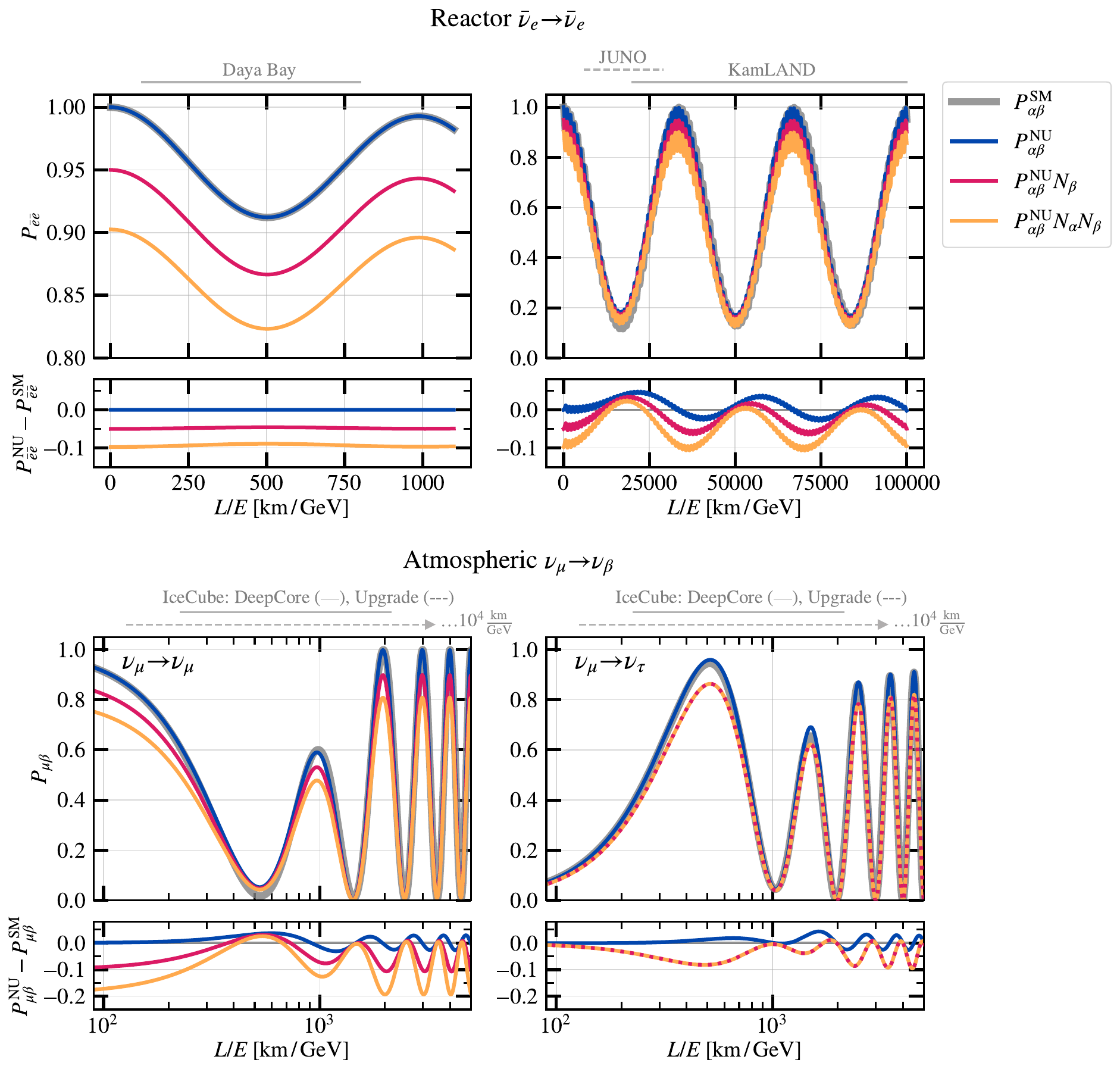}
    \caption{Comparison of the unitary (Standard Model, ``SM'') and the non-unitary (``NU'') oscillation probabilities for the case of $\bar{\nu}_e$ disappearance (top), $\nu_{\mu}$ disappearance (bottom left), and $\nu_{\tau}$ appearance (bottom right), including matter effects. $P_{\alpha\beta}^{\mathrm{NU}}$ corresponds to \cref{eq:nonunitary_osc_probabilities}, while other non-unitary probabilities are scaled by the normalisation factors of $N_{\alpha}$, $N_{\beta}$ for illustration purposes (see text). The approximate ranges of the baseline-to-neutrino-energy ratio ($L/E$) probed by the considered oscillation experiments are shown in gray above each panel. For the atmospheric oscillation probabilities, a fixed baseline of \SI{12742}{km} is assumed. Note that the $P_{\alpha\beta}^{\mathrm{NU}} N_{\beta}$ and $P_{\alpha\beta}^{\mathrm{NU}} N_{\alpha} N_{\beta}$ lines are overlapping in the bottom right panel, since $N_{\alpha} = N_{\mu} = 1$ in this case.}
    \label{fig:oscillation_probs_example}
\end{figure}

To demonstrate the impact of non-unitarity on the oscillation probabilities \eqref{eq:nonunitary_osc_probabilities} compared to the unitary (Standard Model\footnote{Even though neutrino oscillations cannot be explained in the Standard Model without neutrino masses, we use terms ``unitary'' and ``Standard Model'' interchangeably to denote the setup with only 3 neutrino mass states and 3 neutrino flavour states.}) expectations, we focus here on the case of the off-nominal row normalisations. For the purpose of this example, we rescale the row elements of the unitary matrix $U$ (\cref{eq:unitary_matrix_mixing_angles}) such that $N_e = 0.95$ when probing $P_{\bar{e}\bar{e}}$, $N_{\mu} = 0.9$ when probing $P_{\mu\mu}$, and $N_{\tau} = 0.9$ when probing $P_{\mu\tau}$. All other row normalisations, except for the specific one modified in each case, are fixed at 1. We further assume that neutrinos are propagating in a medium with density $\rho = 2.7\,\mathrm{g\,cm^{-3}}$. Our results are shown in \cref{fig:oscillation_probs_example}, both with and without rescaling the probabilities by the factors of $N_{\alpha}, N_{\beta}$ as discussed above. 

From the top left panel of \cref{fig:oscillation_probs_example}, we see that the ``raw'' $P_{\bar{e}\bar{e}}$ probabilities computed from \cref{eq:nonunitary_osc_probabilities} are almost completely unaffected by the off-nominal $N_e$ in the case of a short-baseline experiment such as Daya Bay. This is due to the cancellation of the factors of $N_e$ in the numerator and denominator of \cref{eq:nonunitary_osc_probabilities} and is in agreement with the result obtained by \cite{Ellis:2020hus}. However, the differences with the SM case are clearly visible when rescaling the effective oscillation probability by $N_e$ or $N_e^2$, which is necessary if either the cross section or both the flux and the cross section need to be corrected for non-unitarity. In the remaining three panels, which are representative of the scenarios probed by KamLAND and IceCube-DeepCore, the non-unitary oscillation probability visibly differs from the SM one even if no extra corrections for the cross section ($N_{\beta}$) or the flux ($N_{\alpha}$) are introduced to $P_{\alpha\beta}$. This occurs due to the matter effects, which add an extra $N$-dependent term in the Hamiltonian (see \cref{fig:oscillation_probs_example_vacuum} for comparison with the vacuum case).

Having thereby shown the impact of the normalisation effects on neutrino propagation and detection, we will discuss which choices of the normalisation factors are appropriate for each of the considered experiments in \cref{sec:experiments}.

\section{Experiments}\label{sec:experiments}
\subsection{IceCube}\label{sec:deepcore_upgrade}

\subsubsection{IceCube-DeepCore}\label{sec:deepcore}

The IceCube Neutrino Observatory can precisely probe neutrino oscillations thanks to the detection of atmospheric neutrinos of all three flavours. In the IceCube-DeepCore subarray (later referred to as ``DeepCore'' or ``DC'' in equations and figures), these neutrinos span typical energies from a few GeV to a few hundreds of GeV, with their baselines covering a wide range from \SI{20}{km} (``downgoing'' neutrinos entering the detector from the atmosphere directly) to \SI{12700}{km} (``upgoing'' neutrinos travelling through the entire Earth before reaching IceCube). The events are detected through the Cherenkov light emitted by the electrically charged neutrino interaction products and recorded by the photomultiplier tubes (PMTs) hosted within optical modules (OMs). The time- and space-dependent charge patterns seen by the PMTs enable reconstruction of the neutrino energies and arrival directions, as well as classification of the events based on their topology. The ``track-like'' events include a distinguishable muon track and are representative of $(\nu_{\mu} + \bar{\nu}_{\mu})$-CC interactions. The rest of the events ($(\nu_{e} + \bar{\nu}_{e})$-CC, $(\nu_{\tau} + \bar{\nu}_{\tau})$-CC, and all NC) result in nearly spherical light emission close to the interaction vertex and are classified as ``cascade-like''. The deficit of events in the track-like particle ID (PID) bin therefore provides a probe of atmospheric $\nu_{\mu}$ disappearance \cite{IceCube:2017lak,IceCubeCollaboration:2023wtb}, while the excess of events in the cascade-like PID bin enables $\nu_{\tau}$ appearance measurements \cite{IceCube:2019dqi}. 

In this study, we use the public release of the DeepCore data collected between \SIrange[range-phrase=--,range-units=single]{2012}{2015}{} \cite{IceCube:2019pdr}, which is referred to as ``Sample A'' in \cite{IceCube:2019dqi}. The reconstructed neutrino energies ($E_{\nu}^{\mathrm{reco}}$) span the \SIrange[range-phrase=--,range-units=single]{6}{56}{GeV} range, while  the reconstructed zenith angle $(\theta_{\mathrm{zen}}^{\mathrm{reco}}$) covers the full atmospheric neutrino sky ($\cos\theta_{\mathrm{zen}}^{\mathrm{reco}} \in [-1, 1]$). The release is supplemented by the simulated Monte Carlo neutrino events as well as the estimated muon background, which together pass through the same event selection stages as the data. The PID classification of all events is performed on the basis of the length of the reconstructed track ($\ell_{\mathrm{track}}$) fitted to the recorded light patterns in the detector. All events with $\ell_{\mathrm{track}} > 50\,\mathrm{m}$ are classified as track-like, while the rest fall into the cascade-like PID bin.

While the event-by-event information is provided for both data and neutrino+muon Monte Carlo samples, the statistical analysis is ultimately performed in the binned 3D (reconstructed energy, reconstructed $\cos\theta_{\mathrm{zen}}$, PID) space. The prescriptions for incorporating the impact of the detector systematic uncertainties on the event count in individual bins are also supplied with the data release. This includes the bin count gradients with respect to the variations in the OM efficiency, the ice absorption/scattering coefficients, and the angular dependence of the OM acceptance. The systematic uncertainties related to the atmospheric neutrino flux and the interaction cross section are implemented separately in the public PISA software \cite{IceCube:2018ikn}. The full list of the systematic uncertainties ($\bar{\lambda}_{\mathrm{syst}}$) used for the present analysis follows that of \cite{IceCube:2019dqi}, with the exception of the quasielatic (QE) and the resonance (RES) cross section parameters\footnote{The public DeepCore data release does not include the GENIE \cite{Andreopoulos:2009rq} coefficients necessary to implement the quasielastic and resonance cross section systematic uncertainties. Since the CC-QE (CC-RES) CC events are subdominant at energies $\gtrsim$\SI{6}{GeV} and constitute only $\sim$9\%\,(14\%) of the sample, we find it acceptable to proceed without these two systematics.}. 

Our analysis of the DeepCore data closely follows that of \cite{IceCube:2019dqi} and is based on the procedure devised in \cite{IceCube:2018ikn}. It consists of a staged multiplication of the event weights due to the unoscillated flux, oscillation probability, and the interaction cross section, ultimately followed by binning into the 3D analysis histograms and applying the systematic uncertainty gradients. The expected event rate in each analysis bin therefore depends on the oscillation hypothesis $\bar{\lambda}_{\mathrm{osc}}$ and the systematic parameters $\bar{\lambda}_{\mathrm{syst}}$. For DeepCore, the relevant parameters of $\bar{\lambda}_{\mathrm{osc}}$ include $N_{\mu\{2,3\}}$, $N_{\tau\{2,3\}}$, and the mass splitting $\Delta m^2_{32}$. These parameters directly enter the oscillation probabilities $P_{\mu\mu}(E_{\nu}, L[\cos\theta_{\mathrm{zen}}])$ and $P_{\mu\tau}(E_{\nu}, L[\cos\theta_{\mathrm{zen}}])$ as per \cref{eq:nonunitary_osc_probabilities}. 

Following the discussion of \cref{sec:oscillograms_and_normalisation_effects}, we also need to consider the non-unitarity corrections for the unoscillated fluxes and the interaction cross sections. The nominal atmospheric flux model for the DeepCore analysis is HKKMS-2015 \cite{Honda:2015fha}, which relies on a hadronic interaction model calibrated by the muon spectrometer measurements \cite{Sanuki:2006yd,Honda:2011nf}. We therefore assume that the flux prediction is already contaminated by non-unitarity effects, in particular in the muon sector, and do not apply any additional normalisation corrections related to flux. The nominal cross section model of the analysis, however, relies on Standard Model-based calculations within the GENIE framework \cite{Andreopoulos:2009rq} and therefore requires a normalisation correction $N_{\beta}$ for the final flavour $\beta$ according to \cref{eq:normalisation_corr_cc_xsec,eq:normalisation_corr_nc_xsec}\footnote{An argument was made in \cite{Aloni:2022ebm} that only the deep inelastic scattering (DIS) cross sections need to be corrected for non-unitarity. Since the DIS events constitute the majority ($\gtrsim 70\%$) of the public DeepCore data release, we choose to apply the final-flavour correction to all events regardless of the interaction type. Furthermore, we apply a single correction factor of $N_{\beta}$ to both CC and NC cross sections, since the relative normalisation of the NC events relative to CC is already included as a systematic parameter \cite{IceCube:2019dqi}.}. We absorb this correction factor in the effective oscillation probabilities $\hat{P}_{\alpha\beta}$, such that

\begin{subequations}
\begin{align}
    &\hat{P}_{\mu\mu}^{\mathrm{DC}}(E_{\nu}, L_l\,| \,\bar{\lambda}_{\mathrm{osc}}) = P_{\mu\mu}(E_{\nu}, L_{l}\,| \,\bar{\lambda}_{\mathrm{osc}}) \cdot N_{\mu} = \frac{|(N e^{-iH(\rho_l)L_l} N^{\dag})_{\mu\mu}|^2}{N_{\mu}};\label{eq:norm_corrected_mumu_prob}\\
    &\hat{P}_{\mu\tau}^{\mathrm{DC}}(E_{\nu}, L_l\,| \,\bar{\lambda}_{\mathrm{osc}}) = P_{\mu\tau}(E_{\nu}, L_l\,| \,\bar{\lambda}_{\mathrm{osc}}) \cdot N_{\tau} = \frac{|(N e^{-iH(\rho_l)L_l} N^{\dag})_{\tau\mu}|^2}{N_{\mu}}\label{eq:norm_corrected_mutau_prob}
\end{align}
\label{eq:norm_corrected_deepcore_probs}
\end{subequations}
for the Earth layer $l$ of thickness $L_l$ and constant density $\rho_l$. In this study, we consider a four-layer PREM Earth profile with constant-density layers \cite{Dziewonski:1981xy}. The length of the traversed layers, as well as their number and order, is dependent on the true $\cos\theta_{\mathrm{zen}}$ of the neutrino, which is taken into account in our oscillation probability calculations. The layers are traversed sequentially in their exact order of appearance along the neutrino trajectory for a given $\cos\theta_{\mathrm{zen}}$.

\begin{figure}[htb!]
    \centering
    \includegraphics[width=0.85\textwidth]{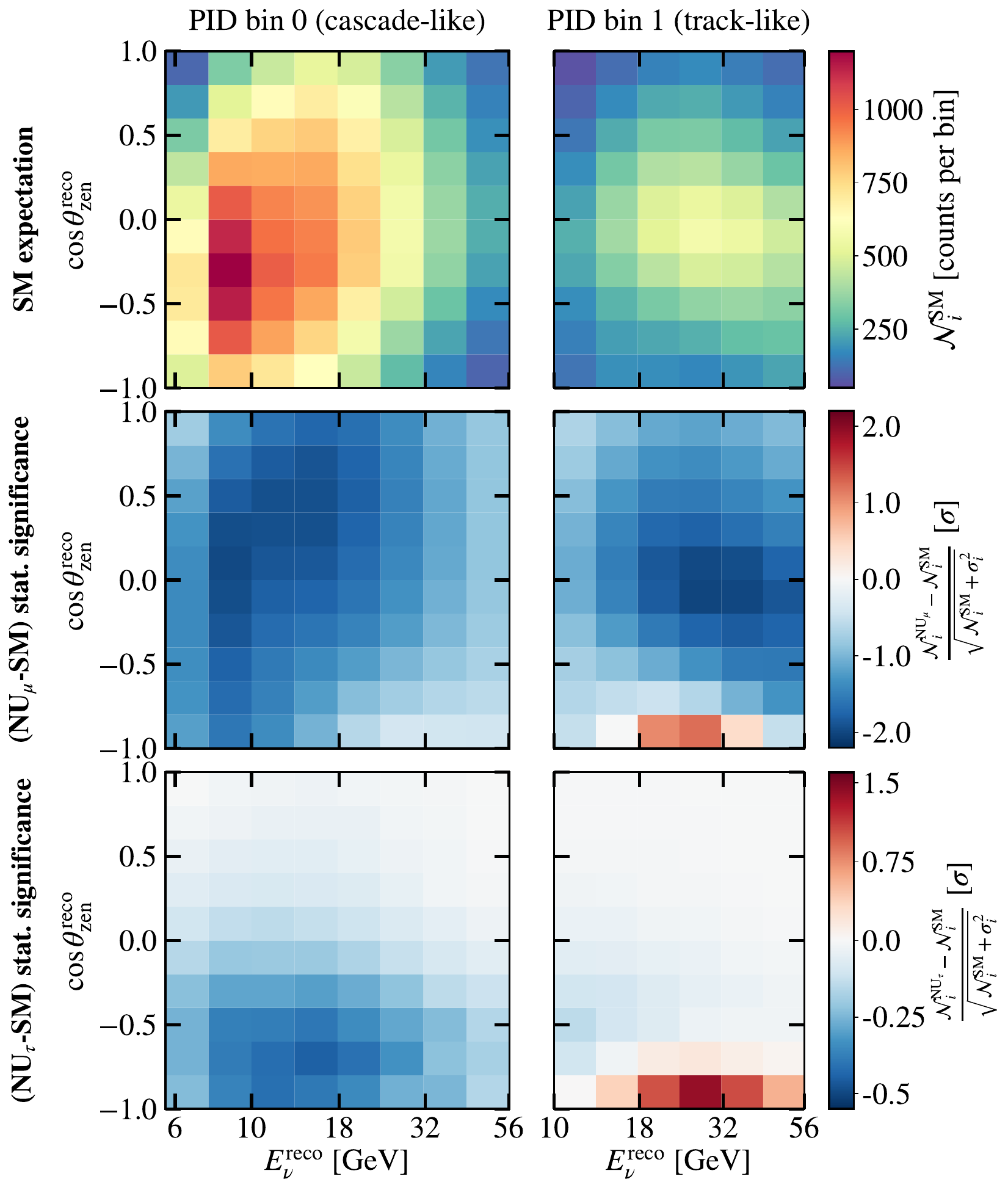}
    \caption{\textit{Top:} the unitary (Standard Model, ``SM'') expectation of the DeepCore event rates in the reconstructed ($E_{\nu}, \cos\theta_{\mathrm{zen}}$) space, assuming the livetime of \SI{2.5}{years}, NuFit 5.2 oscillation parameters \cite{Esteban:2020cvm,NuFit:2022xxx}, and the nominal values of the DeepCore systematic parameters \cite{IceCube:2019dqi,IceCube:2019pdr}. \textit{Middle}: Statistical significance of the non-unitary (``NU'') expectation with $(N_e, N_{\mu}, N_{\tau}) = (1, 0.9, 1)$ compared to the unitary case. \textit{Bottom}: same as the middle panel, but for $(N_e, N_{\mu}, N_{\tau}) = (1, 1, 0.9)$.}
    \label{fig:deepcore_offnominal_numu_templates}
\end{figure}

In \cref{fig:deepcore_offnominal_numu_templates}, we show the impact of the off-nominal muon and tau row normalisations on the DeepCore analysis histograms, where we consider the cases $(N_{\mu} = 0.9, N_{\tau} = 1)$ and $(N_{\mu} = 1, N_{\tau} = 0.9)$ separately. 
As seen from the middle panel of \cref{fig:deepcore_offnominal_numu_templates}, $N_{\mu} < 1$ lowers the expected number of events in both track-like and cascade-like bins, which are dominated by $\nu_{\mu}$-CC events\footnote{Even though $\nu_{\mu}$-CC events contain a true muon track, at low energies it might be classified as a cascade due to the small muon length ($\sim$\SI{4.5}{m/GeV} \cite{Groom:2001kq}) compared to the large DeepCore (\SI{75}{m}) interstring spacing \cite{IceCube:2011ucd}.}. Similarly, the bottom panel shows the deficit of events predominantly occurring in the cascade-like bin for the case of $N_{\tau} < 1$. The excesses of events at $\cos\theta_{\mathrm{zen}}^{\mathrm{reco}} \simeq -1$ in both cases are due to strong matter effects in this region, as it includes neutrinos crossing through the Earth's core (see \cref{fig:deepcore_offnominal_numu_templates_vacuum} for the equivalent figure assuming neutrino propagation in vacuum). The  statistical significances of the event deficits/excesses in the non-unitary mixing case compared to the unitary one are defined as
\begin{equation}
    \Sigma_i = \frac{\mathcal{N}_i^{\mathrm{NU}} - \mathcal{N}_i^{\mathrm{SM}}}{\sqrt{\mathcal{N}_i^{\mathrm{SM}} + \sigma_i^2}}
\label{eq:per_bin_stat_significance}
\end{equation}
for each analysis bin $i$, where $\sigma_i$ is the Monte Carlo uncertainty reflecting the limited simulation statistics. When comparing the observed event rates ($\mathcal{N}^{\mathrm{obs}}$) to the expected ones ($\mathcal{N}^{\mathrm{exp}}$) under the hypothesis  ${\bar{\lambda}}$ = $\{\bar{\lambda}_{\mathrm{osc}},\,\bar{\lambda}_{\mathrm{syst}}\}$, we employ a similar expression for the $\chi^2$ test statistic,
\begin{equation}
    \chi^2_{\mathrm{DC}}(\bar{\lambda}) = \sum_{i=1}^{n_{\mathrm{bins}}} \frac{(\mathcal{N}_i^{\mathrm{obs}}-\mathcal{N}_i^{\mathrm{exp}}(\bar{\lambda}))^2}{\mathcal{N}_i^{\mathrm{exp}}(\bar{\lambda}) + \sigma_i^2}\,+\,\sum_{j=1}^{\mathrm{dim}\,\bar{\lambda}_{\mathrm{syst}}}\left( \frac{s_j - \hat{s}_j}{\sigma_{s_j}} \right)^2.
\label{eq:chi2_deepcore}
\end{equation}
The last term of \cref{eq:chi2_deepcore} is the penalty due to the systematic parameters $s_j \in \bar{\lambda}_{\mathrm{syst}}$, with $\hat{s}_j$ being their nominal values and $\sigma_{s_j}$ -- the widths of the respective Gaussian priors\footnote{If the priors are uniform, no penalty term is added.} \cite{IceCube:2019dqi}. We use the $\chi^2$ from \cref{eq:chi2_deepcore} directly to reproduce the standard oscillation contours for $\nu_{\mu}$ disappearance (see \cref{sec:validation}). For our main Bayesian analysis of the non-unitary mixing, we convert it to the log-likelihood:
\begin{equation}
    \ln{\mathcal{L}_{\mathrm{DC}}}(\bar{\lambda}_{\mathrm{osc}}) = -\frac{1}{2}\chi^2_{\mathrm{DC}},
\label{eq:llh_deepcore}
\end{equation}
assuming that Wilks' theorem \cite{Wilks:1938dza} holds and dropping the $\bar{\lambda}_{\mathrm{osc}}$-independent log-likelihood term for the null hypothesis.

\subsubsection{IceCube-Upgrade}\label{sec:upgrade}

The IceCube-Upgrade (``ICU'' in equations and figures) is an upcoming low-energy enhancement of the IceCube detector, which features 7 additional strings to be deployed within the DeepCore fiducial volume during the polar season \SIrange[range-phrase=--,range-units=single]{2025}{2026}{}. The denser interstring and intermodule configuration compared to DeepCore will enable detection of atmospheric neutrinos with energies as low as $\sim$\SI{1}{GeV} \cite{Ishihara:2019aao}. This will significantly improve the sensitivity of IceCube to both $\nu_{\mu}$ disappearance and $\nu_{\tau}$ appearance. Given 3 years of the IceCube-Upgrade data combined with 12 years of the standard 86-string IceCube configuration, the expected decrease in the width of the ($\sin^2\theta_{23}$, $\Delta m^{2}_{31}$) contours is 20-30\%, while the uncertainty on the $\nu_{\tau}$ normalisation is projected to drop by a factor of two \cite{IceCube:2023ins}.

To incorporate the sensitivity of the IceCube-Upgrade into the future projections for neutrino mixing matrix non-unitarity, we utilize the preliminary public release of the IceCube-Upgrade neutrino Monte Carlo simulation \cite{IceCube:2020umc}. We note that this release corresponds to the sensitivities published in \cite{Ishihara:2019aao} and does not reflect the more recent progress in simulation, event selection, reconstruction, and rejection of random noise and muon background \cite{IceCube:2023ins,IceCube:2022njh}. At the time of the IceCube-Upgrade simulation release, the projected minimum improvement in the $\theta_{\mathrm{zen}}$ reconstruction compared to the DeepCore-only configuration was a factor of $\sim$3. This allows us to bin the $\cos\theta_{\mathrm{zen}}$ dimension of the IceCube-Upgrade expectation templates more finely compared to the corresponding DeepCore templates in \cref{fig:deepcore_offnominal_numu_templates}. We further assume a uniform logarithmic energy binning with 15 bins in the \SIrange[range-phrase=--,range-units=single]{1}{100}{GeV} range. This choice is conservative in the energy range overlapping with that of the three-year DeepCore analysis but includes additional bins at both lower (\SIrange[range-phrase=--,range-units=single]{1}{6}{GeV}) and higher (\SIrange[range-phrase=--,range-units=single]{56}{100}{GeV}) reconstructed energies. The dimensions of the grid were tuned such that the least populated bin had $\gtrsim 20$ counts, which justifies the use of the same test statistic as in \cref{eq:chi2_deepcore,eq:llh_deepcore}.

\begin{figure}[htb!]
    \centering
    \includegraphics[width=0.85\textwidth]{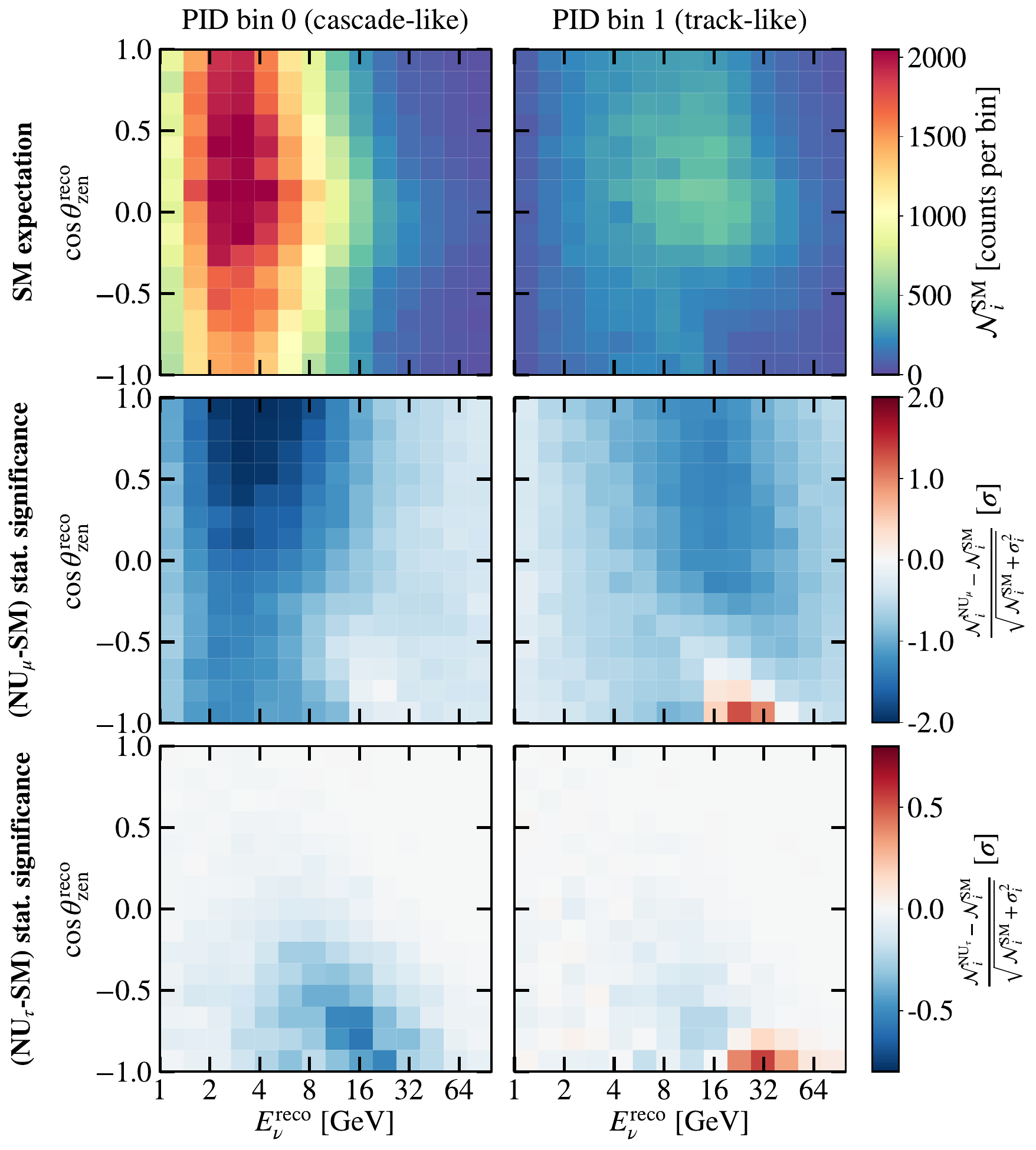}
    \caption{Same as \cref{fig:deepcore_offnominal_numu_templates}, applied to the ICU dataset and binning configuration. The assumed livetime is \SI{3}{years}.} 
    \label{fig:upgrade_offnominal_numu_templates}
\end{figure}

In \cref{fig:upgrade_offnominal_numu_templates}, we show the impact of the off-nominal muon and tau row normalisations ($N_{\mu}$ = 0.9 and $N_{\tau}$ = 0.9, considered separately as in \cref{sec:deepcore}). When comparing these non-unitary expectations to the SM case, we find the same features as in \cref{fig:deepcore_offnominal_numu_templates}, which now provide higher statistical power for placing unitarity constraints due to the finer angular binning and the extended energy range.

For the IceCube-Upgrade analysis, we implement the same atmospheric neutrino flux uncertainties as for the DeepCore analysis (see \cref{sec:deepcore} and \cite{IceCube:2019dqi}), as well as the systematic uncertainty for the relative scaling of the NC cross section compared to CC. The detector systematic uncertainties are not possible to implement for the public IceCube-Upgrade Monte Carlo, as no prescriptions for treating the off-nominal detector systematics (e.g. in the form of the bin count gradients with respect to the systematic parameter variation, as in \cite{IceCube:2019pdr}) are supplied with this release. The per-event GENIE coefficients necessary to implement the QE and the RES cross section systematic uncertainties are similarly not available. We note that the latter will become particularly relevant for the IceCube-Upgrade, as the relative contributions of the CC-QE and CC-RES events to the overall \SIrange[range-phrase=--,range-units=single]{1}{100}{GeV} sample will constitute $\sim$21\% and $\sim$25\%, respectively. Given these limitations, we choose a 3-year livetime for our projections involving the IceCube-Upgrade, which we expect to be statistics-limited. However, we stress that more accurate projections will be possible and necessary in the future once the remaining theoretical and experimental uncertainties are taken into account, along with the recent progress in the technical development of the IceCube-Upgrade simulation \cite{IceCube:2023ins}.

\subsection{Daya Bay}\label{sec:dayabay}

The Daya Bay Reactor Neutrino Experiment (shortly: ``Daya Bay'', later abbreviated as ``DB'' in equations and figures) is probing disappearance of electron antineutrinos from the Daya Bay and the Ling Ao nuclear power plants. The $\bar{\nu}_e$ flux originates from the fission of the $^{235}$U, $^{238}$U, $^{239}$Pu, and $^{241}$Pu isotopes and is detected through the inverse beta decay (IBD) reaction, $\bar{\nu}_{e} + p \to e^{+} + n$, in eight antineutrino detectors (ADs). Four of these detectors are located at the near experimental halls, EH1 and EH2, and the remaining four are placed at the far experimental hall, EH3. The distances between the reactor cores and the far detectors range between \SI{1.5}{km} and \SI{1.9}{km}, making Daya Bay a \textit{short-baseline} detector. The IBD reaction products ionize the liquid scintillator filling the ADs, which leads to the emission of photons detected by the PMTs in each detector. This enables the reconstruction of the deposited ``prompt energy'':
\begin{equation}
    E_{\mathrm{prompt}} \approx E_{\bar{\nu}} + m_p - (m_n + m_e + T_n) + 2 E_{\gamma},
\label{eq:neutrino_to_prompt_energy_full}
\end{equation}
where $E_{\bar{\nu}}$ is the energy of the incoming antineutrino; $m_p$, $m_n$, and $m_e$ are the masses of the proton, neutron, and electron, respectively; $T_n$ is the kinetic energy of the neutron; and $2E_{\gamma} = 2 \times 0.511\,\mathrm{MeV}$ is the energy of the two photons emitted as the result of the $e^{+}e^{-}$ annihilation in the scintillator medium. The typical range of $T_n$ is $\mathcal{O}(10\,\mathrm{keV})$ \cite{DayaBay:2016ssb} and is therefore neglected. This leads to the following approximation: 
\begin{equation}
    E_{\mathrm{prompt}} \approx E_{\bar{\nu}} - 0.782\,\mathrm{MeV},
\label{eq:positron_energy}
\end{equation}
which we adopt in this paper\footnote{We subsequently assume that the 26 prompt energy bins used in the official analyses by the Daya Bay Collaboration \cite{DayaBay:2018yms,DayaBay:2022orm} have a 1:1 correspondence with the 26 energy bins of the IBD neutrino spectrum from \cite{DayaBay:2016ssb}, which is an approximation to the complete energy response matrix of the detector (see e.g. \cite{DayaBay:2016ssb,DayaBay:2021dqj}).}. 
To probe the electron row of the mixing matrix $N$ entering \cref{eq:nonunitary_osc_probabilities}, we employ a method similar to the Analysis Method B from \cite{DayaBay:2016ggj}, as was done in the most recent analysis by the Daya Bay Collaboration at the time of writing \cite{DayaBay:2022orm}. 

To predict the event rates at the far hall EH3, Method B relies on convolution of the reactor $\bar{\nu}_e$ flux model, the standard IBD cross section \cite{Vogel:1999zy}, the oscillation probabilities $P_{\bar{e}\bar{e}}(\bar{\lambda}_{\mathrm{osc}})$, and the detector response. Here,  $\bar{\lambda}_{\mathrm{osc}}$ denotes the parameters driving the $\bar{\nu}_e$ survival probability, which for Daya Bay are the electron row elements $N_{ei}$ and the mass splitting $\Delta m^2_{32}$. The reactor flux model is effectively constrained by the near hall measurements \cite{DayaBay:2016ggj}, and therefore is already contaminated by any non-unitarity effects. However, the IBD cross section used in this analysis is based on a calculation assuming the Standard Model and thus requires an extra correction factor of $N_e \equiv (NN^{\dag})_{ee}$. We include this factor in the effective oscillation probabilities $\hat{P}_{\bar{e}\bar{e}}(\bar{\lambda}_{\mathrm{osc}})$, which become
\begin{equation}
    \hat{P}_{\bar{e}\bar{e}}^{\mathrm{DB}}(E_{\bar{\nu}_e}, L_r\,| \,\bar{\lambda}_{\mathrm{osc}}) = P_{\bar{e}\bar{e}}(E_{\bar{\nu}_e}, L_r\,| \,\bar{\lambda}_{\mathrm{osc}}) \cdot N_e = \frac{|(N e^{-iHL_r} N^{\dag})_{ee}|^2}{N_e}
\label{eq:effective_osc_prob_dayabay}
\end{equation}
for a fixed baseline $L_r$ connecting the far hall with one of the six reactors $r$. We further perform averaging over reactor baselines  as follows:
\begin{equation}
    \langle\hat{P}_{\bar{e}\bar{e}}^{\mathrm{DB}}(E_{\bar{\nu}_e}\,| \,\bar{\lambda}_{\mathrm{osc}})\rangle_L = \Big[\sum_r \frac{1}{L_r^2} \hat{P}_{\bar{e}\bar{e}}^{\mathrm{DB}}(E_{\bar{\nu}_e}, L_r\,| \,\bar{\lambda}_{\mathrm{osc}})\Big] \cdot \Big[ \sum_r \frac{1}{L_r^2} \Big]^{-1},
\label{eq:oscillation_baseline_averaging}
\end{equation}
such that the average oscillation probability $\langle\hat{P}_{\bar{e}\bar{e}}^{\mathrm{DB}}(E_{\bar{\nu}_e}\,| \,\bar{\lambda}_{\mathrm{osc}})\rangle_L$ only depends on neutrino energy and the oscillation parameters.

In this study, we analyze the IBD candidate selection from 3158 days of Daya Bay data \cite{DayaBay:2022orm}. Together with the release of this dataset, the Daya Bay Collaboration provides the event rates predicted for EH3 with the best-fit oscillation parameters: $\Delta m_{32} = (2.466 \pm 0.060) \cdot 10^{-3}\,\mathrm{eV^2}$ (under normal mass ordering) and $\sin^2{2\theta_{13}} = 0.0851 \pm 0.0024$. We unfold this best-fit oscillation prediction to compute the expected unoscillated event rates at EH3:
\begin{equation}
    \mathcal{N}^{\mathrm{exp,\,no\,osc}}_{\mathrm{EH3}} (E_{\mathrm{prompt}}) = \frac{\mathcal{N}^{\mathrm{exp}}_{\mathrm{EH3}}(E_{\mathrm{prompt}}, \bar{\lambda}_{\mathrm{osc}}^{\mathrm{best\,fit}})}{\langle\hat{P}_{\bar{e}\bar{e}}^{\mathrm{DB}}(E_{\bar{\nu}_e}\,| \,\bar{\lambda}_{\mathrm{osc}}^{\mathrm{best\,fit}})\rangle_{L}},
\end{equation}
where $\bar{\lambda}_{\mathrm{osc}}^{\mathrm{best\,fit}}$ assumes unitary mixing ($NN^{\dag} = N^{\dag}N = \mathbb{I}$). We then use this unoscillated prediction to compute the expected spectrum $\mathcal{N}^{\mathrm{\,exp}}_{\mathrm{EH3}}$ under the tested non-unitary mixing hypothesis $\bar{\lambda}_{\mathrm{osc}}$:
\begin{equation}
    \mathcal{N}^{\mathrm{exp}}_{\mathrm{EH3}} (E_{\mathrm{prompt}}, \bar{\lambda}_{\mathrm{osc}}) = {\mathcal{N}^{\mathrm{exp,\,no\,osc}}_{\mathrm{EH3}}(E_{\mathrm{prompt}}) \cdot \langle\hat{P}_{\bar{e}\bar{e}}^{\mathrm{DB}}(E_{\bar{\nu}_e}\,| \,\bar{\lambda}_{\mathrm{osc}})\rangle_L}.
\end{equation}

Finally, the expected spectrum $\mathcal{N}^{\mathrm{exp}}_{\mathrm{EH3}}$ is  compared to the observed spectrum $\mathcal{N}^{\mathrm{obs}}_{\mathrm{EH3}}$ at EH3. To optimize for the parameters of $\bar{\lambda}_{\mathrm{osc}}$, the following $\chi^2$ test statistic is constructed\footnote{We note that the original version of the Analysis Method B from \cite{DayaBay:2016ggj} relies on profiling of the systematic parameters with  respective penalty terms added to the $\chi^2$. However, since we do not have access to the Daya Bay simulation chain and implementaton of the individual systematics, we resort to the covariance matrix method instead.}:
\begin{equation}
    \chi^2_{\mathrm{DB}}(\bar{\lambda}_{\mathrm{osc}}) = (\mathcal{N}^{\mathrm{obs}}_{\mathrm{EH3}} - \mathcal{N}^{\mathrm{exp}}_{\mathrm{EH3}})^{\top} V^{-1}_{\mathrm{tot}} (\mathcal{N}^{\mathrm{obs}}_{\mathrm{EH3}} - \mathcal{N}^{\mathrm{exp}}_{\mathrm{EH3}}),
\label{eq:chi2_daya_bay}
\end{equation}
where $V_{\mathrm{tot}}$ is the total covariance matrix consisting of the statistical and the systematic components, $V_{\mathrm{tot}} = V_{\mathrm{stat}} + V_{\mathrm{syst}}$. The statistical component of $V$ is calculated as a diagonal matrix with $V_{\mathrm{stat},\,ii} = \mathcal{N}^{\mathrm{\,exp}}_{\mathrm{EH3}, i}$, where index $i$ runs over the prompt energy bins $E_{{\mathrm{prompt},\,i}}$. The estimation of the systematic component follows the prescription from \cite{DayaBay:2016ssb} and is detailed in \cref{sec:db_covariance}\footnote{The ultimate impact of the non-zero $V_{\mathrm{syst}}$ on the measured electron row elements is shown in \cref{fig:dayabay_systematic_impact}.}. As in \cref{sec:deepcore_upgrade}, the $\chi^2$ from \cref{eq:chi2_daya_bay} is used to reproduce the three-flavour Daya Bay contours in \cref{sec:validation}, and converted to the log-likelihood $\ln\mathcal{L}_{\mathrm{DB}}$ analogously to \cref{eq:llh_deepcore} for the Bayesian analysis of the non-unitary mixing.

\begin{figure}[htb!]
    \centering
    \includegraphics[width=0.6\textwidth]{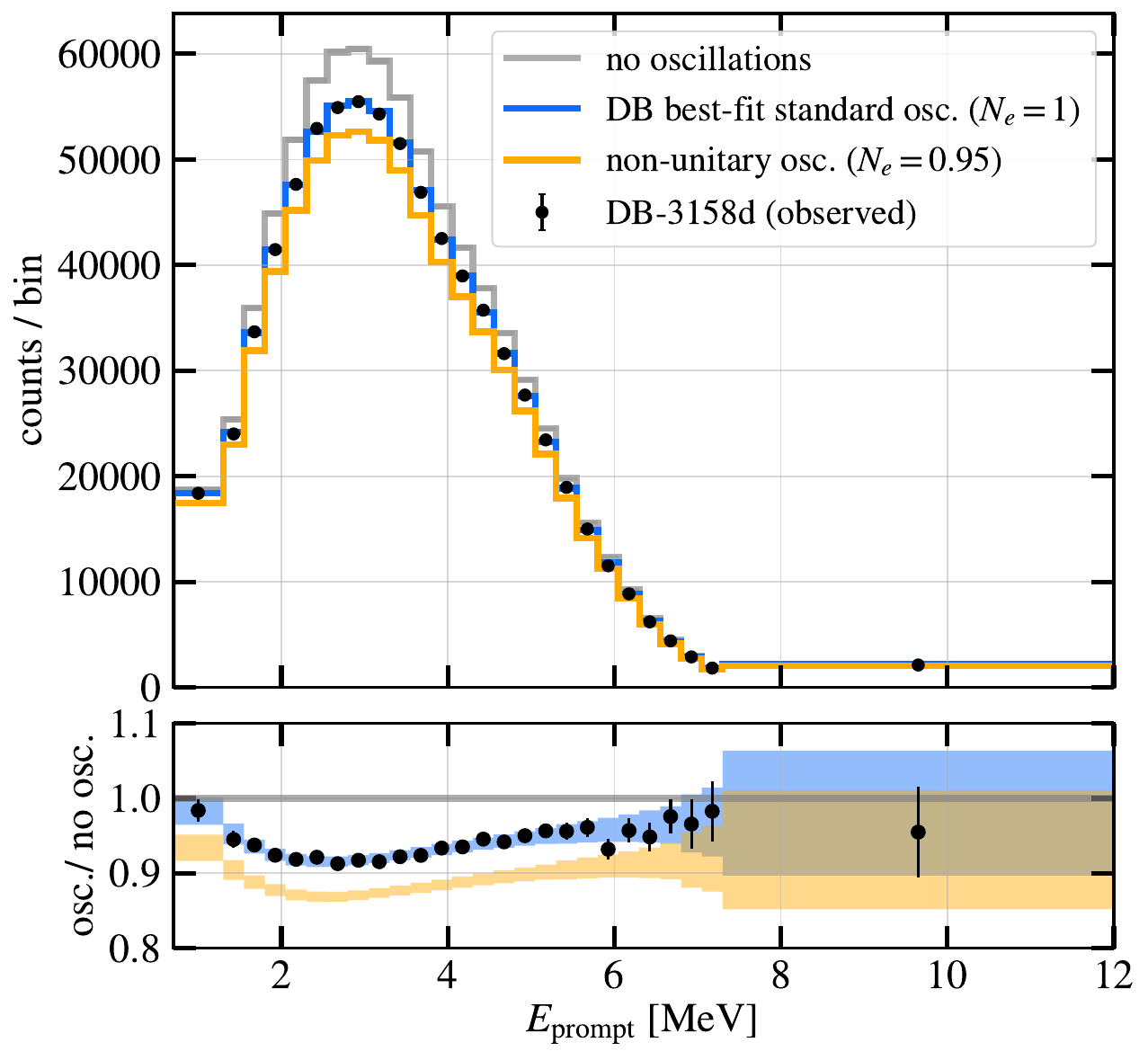}
    \caption{\textit{Top}: Daya Bay prompt energy spectrum predicted at the far hall EH3 for the case of no oscillations (gray), Daya Bay best-fit three-flavour oscillations (blue) \cite{DayaBay:2022orm}, and non-unitary oscillations with normalisation of the electron row $N_e = 0.95$ (yellow). The black dots represent the spectrum observed by Daya Bay at EH3 in 3158 days. \textit{Bottom}: ratios of the expected or observed spectra including oscillations to the no-oscillation prediction. The color scheme follows that of the top panel. The errorbars represent the statistical uncertainty for data and the square-roots of the diagonal elements of the covariance matrix ($\sqrt{V_{ii}}$) for expectations.}
    \label{fig:daya_bay_histogram}
\end{figure}

In \cref{fig:daya_bay_histogram}, we show the predicted prompt energy spectrum at EH3 assuming no oscillations, $\mathcal{N}^{\mathrm{exp,\,no\,osc.}}_{\mathrm{EH3}}$; the actual spectrum measured by Daya Bay in 3158 days of livetime, $\mathcal{N}^{\mathrm{obs}}_{\mathrm{EH3}}$; and the corresponding best-fit spectrum, $\mathcal{N}^{\mathrm{exp}}_{\mathrm{EH3}}(\bar{\lambda}_{\mathrm{osc}}^{\mathrm{best\,fit}})$. All of the spectra are background-subtracted and summed over the data-taking periods. To illustrate the effect of non-unitarity on the predicted event rate, we scale the electron row of the mixing matrix $N$ such that the normalisation $N_e = 0.95$, i.e., 5\% lower than the unitary expectation. 
We see from \cref{fig:daya_bay_histogram} that the deviation from the measured Daya Bay spectrum in the non-unitary case is highly non-statistical, which suggests that the constraint on the electron row normalisation that could be put with Daya Bay will be much stronger than 5\%. We note that this is the result of the IBD cross section correction by one factor of $N_e$. This factor would cancel out if one were to use an analysis method based on computing the far/near detector event ratios, such as the Method A of Ref.\,\cite{DayaBay:2016ggj}, resulting in the loss of sensitivity to the electron row normalisation with Daya Bay alone.

\subsection{KamLAND}\label{sec:kamland}

The concept of the Kamioka Liquid Scintillator Antineutrino Detector (``KamLAND'', later abbreviated as ``KL'' in equations and figures) is similar to that of Daya Bay, and we will reuse some of the concepts from \cref{sec:dayabay} for brevity. KamLAND measures electron antineutrinos from nuclear reactors via the IBD process in a single tank filled with liquid scintillator, and the light emitted post-interaction due to the scintillation and Cherenkov radiation processes is collected by PMTs. The main difference compared to Daya Bay is that there are more than 50 (57 in this study) contributing nuclear reactors located at much larger distances to the detector, with the effective flux-averaged distance $L_0 \simeq $\SI{180}{km}. This means that KamLAND can probe $\sim$100 times smaller mass splittings compared to Daya Bay, which makes it sensitive to $\Delta m^{2}_{21}$ and $N_{e\{1,2\}}$ (alternatively, $\theta_{12}$ in the mixing angle parameterisation). Furthermore, due to both long baselines and the depth at which the KamLAND detector is located (\SI{2700}{m} w.e.), matter effects start playing a role in neutrino oscillations and have to be taken into account explicitly in the Hamiltonian of \cref{eq:matter_hamiltonian_mass_basis}. 

In this study, we analyze the selection of IBD candidates collected by KamLAND over 7 years (\SIrange[range-phrase=--,range-units=single]{2002}{2009}{}), as reported in \cite{KamLAND:2010fvi}. From the referenced publication, we extract the no-oscillation prediction for the binned prompt energy spectrum at the detector and subtract the best-fit background distributions. We assume that this energy spectrum directly corresponds to the energy deposited by the IBD interaction products, i.e., we do not apply any deposited-to-visible energy conversions due to non-linearity of the detector response\footnote{We investigated this option in our analysis, applying the $E_{\mathrm{vis}} (E_{\mathrm{dep}})$ correction as derived in \cite{Enomoto:2005xxx}, but found nearly no change in our reproduction of the standard oscillation results.}. However, we take into account the 1.9\% uncertainty on the energy scale determination ($\epsilon_{E}$) and the 4.1\% bin-to-bin correlated uncertainty on the event rate ($\epsilon_{\mathcal{N}}$), which are the dominant systematic uncertainties for the determination of $\Delta m^2_{21}$ and $\theta_{12}$ \cite{KamLAND:2008dgz,KamLAND:2010fvi}\footnote{The impact of these uncertainties on the measured electron row elements is shown in \cref{fig:kamland_systematic_impact}.}.

We compute the expected energy spectrum under the oscillation hypothesis $\bar{\lambda}_{\mathrm{osc}}$ as 
\begin{equation}
    \mathcal{N}_{\mathrm{exp}}(E_{\mathrm{prompt}},\bar{\lambda}_{\mathrm{osc}}, \bar{\lambda}_{\mathrm{syst}}) = \mathcal{N}_{\mathrm{exp,\,no\,osc}}(E_{\mathrm{prompt}}, \bar{\lambda}_{\mathrm{syst}}) \,\langle\hat{P}_{\bar{e}\bar{e}}^{\mathrm{KL}}(E_{\bar{\nu}_e}\,| \,\bar{\lambda}_{\mathrm{osc}})\rangle_L,
\end{equation}
where $\langle\,\rangle_{L}$ denotes averaging over the baselines as in \cref{eq:oscillation_baseline_averaging} and $\bar{\lambda}_{\mathrm{syst}} = (\epsilon_{E}, \epsilon_{\mathcal{N}})$ are the systematic parameters. To arrive at the effective survival probability $\hat{P}_{\bar{e}\bar{e}}^{\mathrm{KL}}$ for KamLAND, the standard oscillation probability is multiplied by two factors of $N_e$. This is needed to correct for non-unitarity in both the IBD cross section and the reactor flux, which are based on calculations assuming the Standard Model rather than calibration at the near detector \cite{Vogel:1999zy,Detwiler:2005xxx}. This removes the factors of $N_e$ from the denominator of \cref{eq:nonunitary_osc_probabilities} and yields
\begin{equation}
        \hat{P}_{\bar{e}\bar{e}}^{\mathrm{KL}}(E_{\bar{\nu}_e}, L_r\,| \,\bar{\lambda}_{\mathrm{osc}}) = |(N e^{-iHL_r} N^{\dag})_{ee}|^2
\end{equation}
for a fixed baseline $L_r$.

Due to the relatively low numbers of counts in the KamLAND spectra, we are using the binned Poisson likelihood\footnote{We note that this treatment differs from the unbinned and time-dependent likelihood approach officially used by the KamLAND collaboration \cite{KamLAND:2008dgz,KamLAND:2010fvi}; however, since we do not have access to the event-by-event information from KamLAND, we find this approach to be optimal given the public information available.} to quantify the agreement of the observed data with the expected spectrum under a given oscillation hypothesis:
\begin{equation}
    \ln{\mathcal{L}_{\mathrm{KL}}(\bar{\lambda}_{\mathrm{osc}}, \bar{\lambda}_{\mathrm{syst}})} = \mathcal{N}_{\mathrm{obs}} \cdot \ln{\left[\mathcal{N}_{\mathrm{exp}}(\bar{\lambda}_{\mathrm{osc}}, \bar{\lambda}_{\mathrm{syst}})\right]} - \mathcal{N}_{\mathrm{exp}}(\bar{\lambda}_{\mathrm{osc}}, \bar{\lambda}_{\mathrm{syst}}) - \ln{\left[\mathcal{N}_{\mathrm{obs}}!\right]}. 
\end{equation}
The penalty terms due to $\bar{\lambda}_{\mathrm{syst}}$ are added in the same way as in \cref{eq:chi2_deepcore}.

\begin{figure}[htb!]
    \centering
    \includegraphics[width=0.6\textwidth]{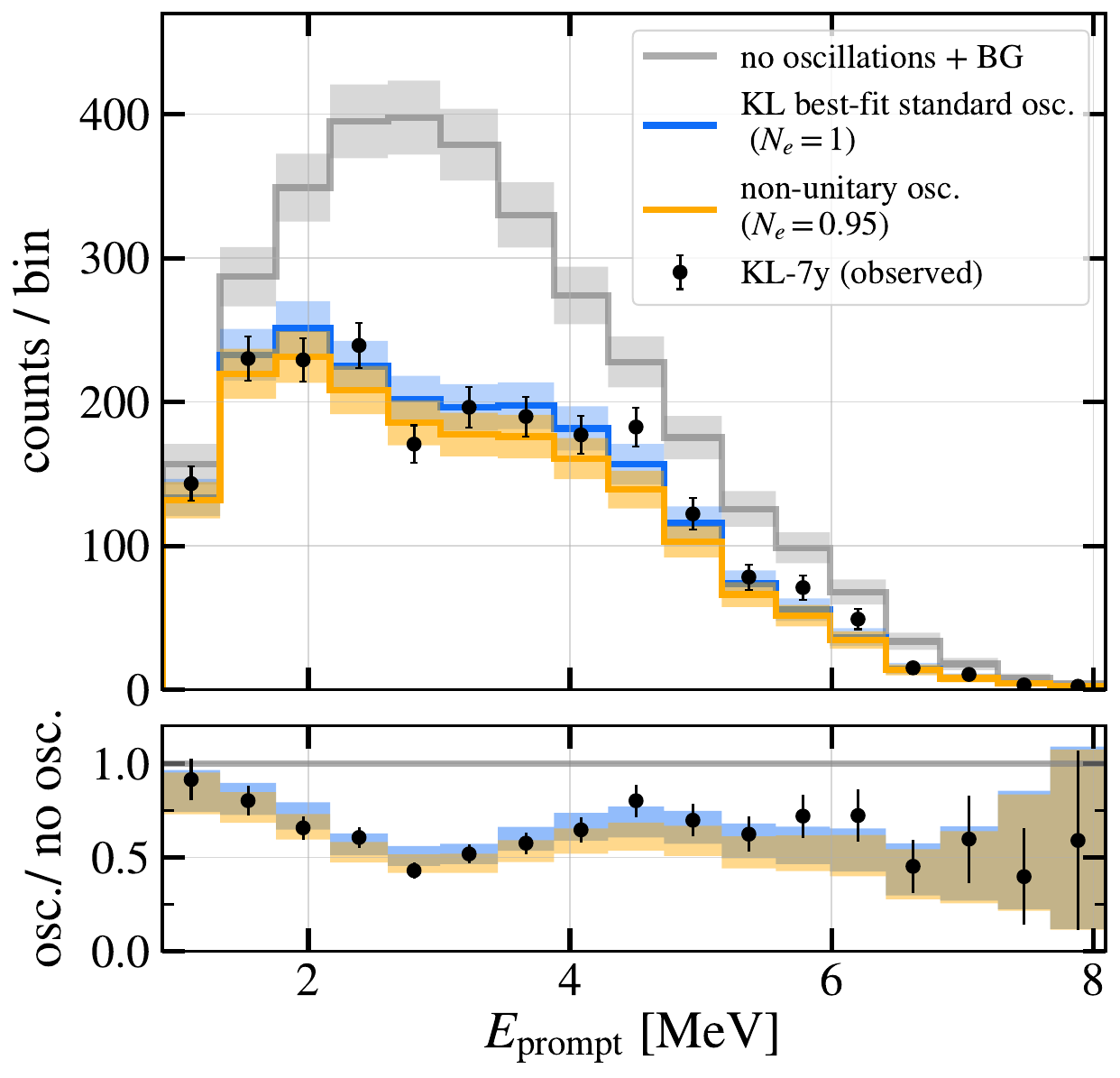}
    \caption{\textit{Top}: KamLAND prompt energy spectrum predicted for the case of no oscillations (gray), KamLAND best-fit two-flavour oscillations (blue) \cite{KamLAND:2010fvi}, and non-unitary oscillations with normalisation of the electron row $N_e = 0.95$ (yellow). The black dots represent the spectrum observed by KamLAND between 2002-2009. \textit{Bottom panel}: same as \cref{fig:daya_bay_histogram}. The errorbars represent the statistical uncertainty for data and the total (statistical + systematic) uncertainty on the event rate for expectations.}
    \label{fig:kamland_histogram}
\end{figure}

In \Cref{fig:kamland_histogram}, we show the predicted prompt energy spectrum assuming no oscillations, the actual spectrum measured by KamLAND over 7 years, and the spectrum expected at the KamLAND best-fit point assuming $\theta_{13} = 0^{\circ}$ ($\Delta m^2_{21} = 7.5 \cdot 10^{-5}\,\mathrm{eV}^2$, $\tan^2\theta_{12} = 0.492$) \cite{KamLAND:2010fvi}. As was done in \cref{fig:daya_bay_histogram}, we also provide the corresponding prediction under the assumption of non-unitary mixing, where we rescale all of the elements of the electron row such that $N_e = 0.95$. 
We see that due to limited statistics, KamLAND is not nearly as sensitive to the electron row normalisation as Daya Bay, and therefore only a weak improvement to the Daya Bay constraint on $N_e$ is expected from inclusion of KamLAND. However, KamLAND provides an additional handle on the $N_{e1}$ and the $N_{e2}$ elements of the mixing matrix, which enter the KamLAND oscillation probability in a different combination ($|N_{e1}|^2\,|N_{e2}|^2$) than that of Daya Bay ($|N_{e1}|^2 + |N_{e2}|^2$) \cite{Ellis:2020hus}. Thus, we include KamLAND into the analysis for the purpose of constraining $|N_{e1}|$ and $|N_{e2}|$ individually. At the same time, we note that this cannot be done completely unambibuously without the solar neutrino oscillation data (such as that from SNO \cite{SNO:2011hxd}), which helps break the degeneracy in the octant of $\theta_{12}$. We therefore use prior information from the solar neutrino measurements that $\theta_{12}$ lies in the lower octant, such that $|N_{e1}| > |N_{e2}|$, which helps avoid the bimodality in our ultimate matrix element posteriors. 

\subsection{JUNO}\label{sec:juno}

The Jiangmen Underground Neutrino Observatory (``JUNO'') is an upcoming multipurpose neutrino experiment, which is undergoing installation in Southern China at the time of writing \cite{Guo:2024tD,JUNO:2022mxj}. It will measure, in particular, the disappearance of reactor $\bar{\nu}_e$ from two nuclear power plants located at \SI{52.5}{km} distance each from a liquid scintillator detector. The combination of this baseline length with the typical energy range of the IBD spectrum ($\sim$\SIrange[range-phrase=--,range-units=single]{1.8}{9}{MeV}) results in a sufficiently large $L/E$ reach suitable for measuring the slow oscillations due to $\sin^2\theta_{12}$ and $\Delta m^2_{21}$. At the same time, the energy deposited by the IBD interaction products will be measured in JUNO with very high resolution ($\sim$3\% at \SI{1}{MeV} using the main JUNO PMT system, LPMT \cite{JUNO:2022mxj}). This will allow for simultaneous measurements of $\Delta m^2_{31}$ and $\sin^2\theta_{13}$, which induce fast oscillations. 

The JUNO Collaboration provides the prediction for the unoscillated event rates $\mathcal{N}_{\mathrm{exp,\,no\,osc}}$ as a function of true neutrino energy, which results from a convolution of the reactor fluxes with the IBD cross section \cite{JUNO:2022mxj}. Since the assumed cross section is following the Standard Model-based calculation by \cite{Vogel:1999zy}, we correct it by the normalisation $N_e$ as per \cref{eq:normalisation_corr_cc_xsec}. This correction factor is absorbed into the effective oscillation probability $\hat{P}_{ee}$ as in \cref{eq:effective_osc_prob_dayabay}. The reactor flux, on the other hand, does not require a non-unitarity correction, since its calibration at the satellite TAO detector is envisioned \cite{JUNO:2022mxj,JUNO:2020ijm}. 

To represent the expected oscillated event rates $\mathcal{N}_{\mathrm{exp,\,no\,osc}} (E_{\nu}) \cdot \hat{P}_{ee}(L, E_{\nu})$ as a function of the visible energy $E_{\mathrm{vis}}$ in the detector, we apply the shift from $E_{\nu}$ to the prompt (positron) energy as per \cref{eq:positron_energy}, which we further correct according to the nonlinear response of the LPMT system and smear according to its expected resolution. Following \cite{JUNO:2022mxj}, we use a \SI{20}{keV} bin width for the resulting visible energy range. For the final oscillated spectrum expectation, we assume a 2.2\% flux uncertainty ($\epsilon_{\mathcal{N}}$) and a 1\% detection uncertainty ($\epsilon_{\mathrm{eff}}$), both bin-to-bin correlated. We further include the geoneutrino background, which is the dominant source of background in JUNO \cite{JUNO:2022mxj}, as well as an associated 30\% uncertainty ($\epsilon_{\mathrm{geo}}$). We do not take into account other sources of background or spectrum shape uncertainties in this analysis. The $\epsilon_{\mathcal{N}}$, $\epsilon_{\mathrm{eff}}$, and $\epsilon_{\mathrm{geo}}$ systematic parameters acquire respective penalty terms in the $\chi^2_{\mathrm{JUNO}}$ test statistic. The latter is equivalent to \cref{eq:chi2_deepcore}, modulo the Monte Carlo simulation uncertainty $\sigma_i$, which only applies to DeepCore. As in all other experiments for which the $\chi^2$ metric is defined, we convert it to the log-likelihood $\ln\mathcal{L}_{\mathrm{JUNO}}$ assuming Wilks' theorem.

\begin{figure}[htb!]
    \centering
    \includegraphics[width=0.6\textwidth]{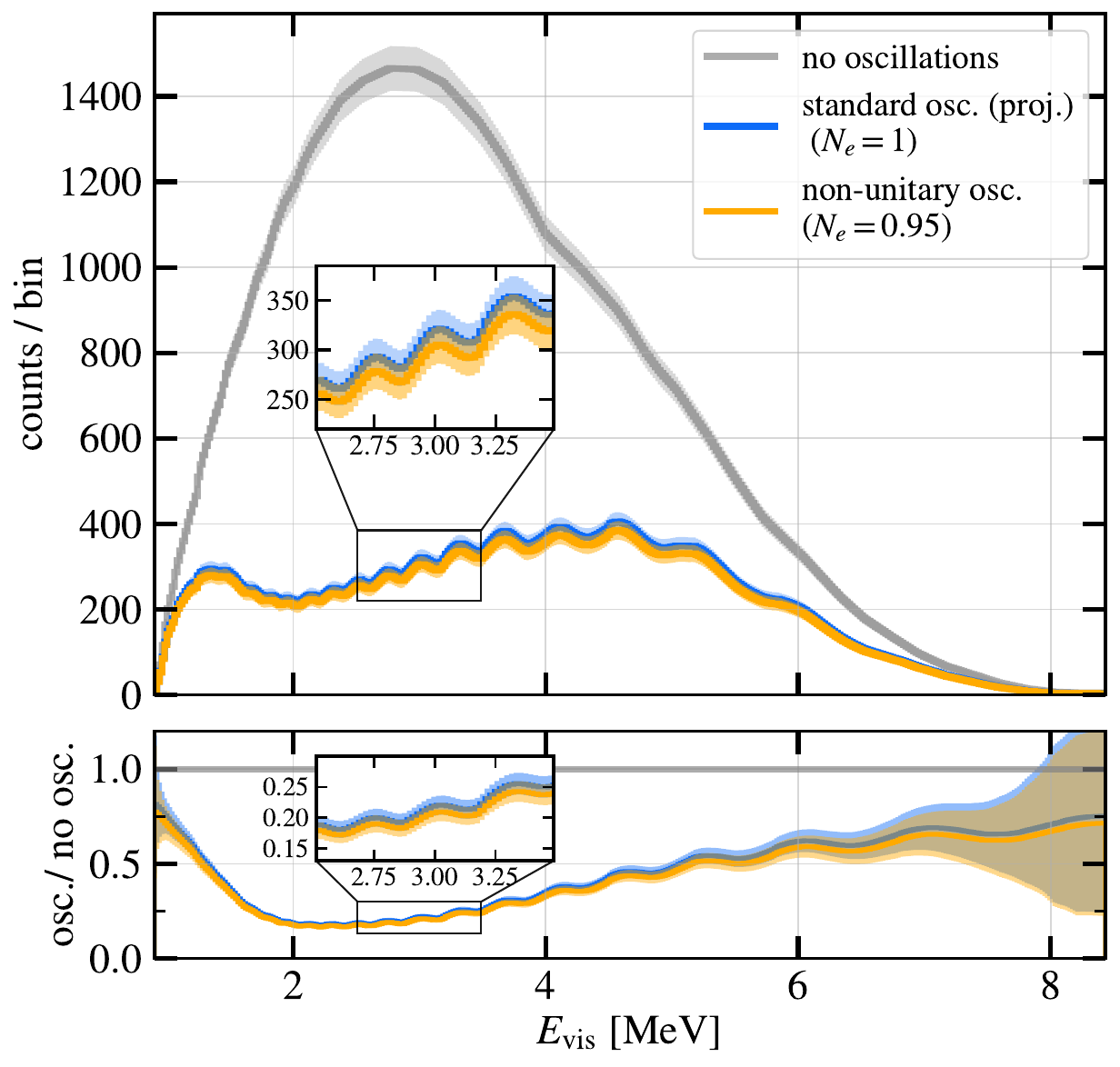}
    \caption{\textit{Top}: JUNO visible energy spectrum predicted for the case of no oscillations (gray), best-fit oscillations with parameters from \cite{JUNO:2022mxj}, and non-unitary oscillations with normalisation of the electron row $N_e = 0.95$ (yellow). \textit{Bottom}: same as \cref{fig:daya_bay_histogram}. The errorbars represent the total (statistical + systematic) uncertainty on the expected event rate. The backgrounds are not included in this figure. The assumed livetime is 6 years.}  
    \label{fig:juno_histogram}
\end{figure}

In \cref{fig:juno_histogram}, we show our predictions for the visible energy spectra in JUNO after 6 years of exposure for the same physics cases as in \cref{fig:daya_bay_histogram,fig:kamland_histogram}. By comparing the spectrum generated under the standard three-flavour scenario to the non-unitary mixing case with $N_e = 0.95$, we can expect that JUNO will similarly have the capability to constrain the electron row normalisation but not improve significantly on the Daya Bay-only constraints.

\section{Validation: Reproduction of the three-flavour oscillation results}\label{sec:validation}

Prior to fitting for the elements of the non-unitary mixing matrix $N$, we validate our setup of the experiments, systematic uncertainties, and test statistics by reproducing the standard three-flavour oscillation results. In particular, we construct the following confidence level contours: ($\sin^2{2\theta_{13}}$, $\Delta m^2_{32}$) for Daya Bay; ($\tan^2{\theta_{12}}$, $\Delta m^2_{21}$) for KamLAND; ($\sin^2{\theta_{12}}$, $\Delta m^2_{21}$) for JUNO; and ($\sin^2{\theta_{23}}$, $\Delta m^2_{32}$) for IceCube-DeepCore and IceCube-Upgrade. The contours are derived from the frequentist $\chi^2$ scans across the relevant 2D parameter spaces and compared to the official results from the respective collaborations \cite{DayaBay:2022orm,KamLAND:2008dgz,IceCube:2019dqi,JUNO:2022mxj,Ishihara:2019aao}. The solar mass splitting and mixing angle are fixed at $\Delta m^2_{21} = 7.53\,\cdot 10^{-5}\,\mathrm{eV^2}\,\,(7.5\,\cdot 10^{-5}\,\mathrm{eV^2})$ and $\theta_{12} = 33.48^{\circ}\,(33.64^{\circ})$ for the Daya Bay (DeepCore) analysis. For future projections with JUNO and IceCube-Upgrade, the injected true parameters are ($\sin^2{\theta_{12}}$, $\Delta m^2_{21}$) = (0.307, $7.53 \cdot 10^{-5}\,\mathrm{eV^2}$) and ($\sin^2{\theta_{23}}$, $\Delta m^2_{32}$) = (0.51, $2.31 \cdot 10^{-3}\,\mathrm{eV^2}$), respectively, following \cite{JUNO:2022mxj,IceCube:2019dqi}. The DeepCore systematic parameters $\bar{\lambda}_{\mathrm{syst}}$ \cite{IceCube:2019dqi} are profiled over, i.e., fitted by minimizing $\chi^2_{\mathrm{DC}} (\bar{\lambda}_{\mathrm{osc}},\,\bar{\lambda}_{\mathrm{syst}})$ at a given scan point. The same treatment applies to the simplified KamLAND and JUNO systematic parameters (see \cref{sec:kamland,sec:juno}). To reproduce the KamLAND results, we additionally fix $\theta_{13} = 0^{\circ}$ as was done in the reference study \cite{KamLAND:2008dgz}.

\begin{figure}[htb!]
    \centering
    \includegraphics[width=\textwidth]{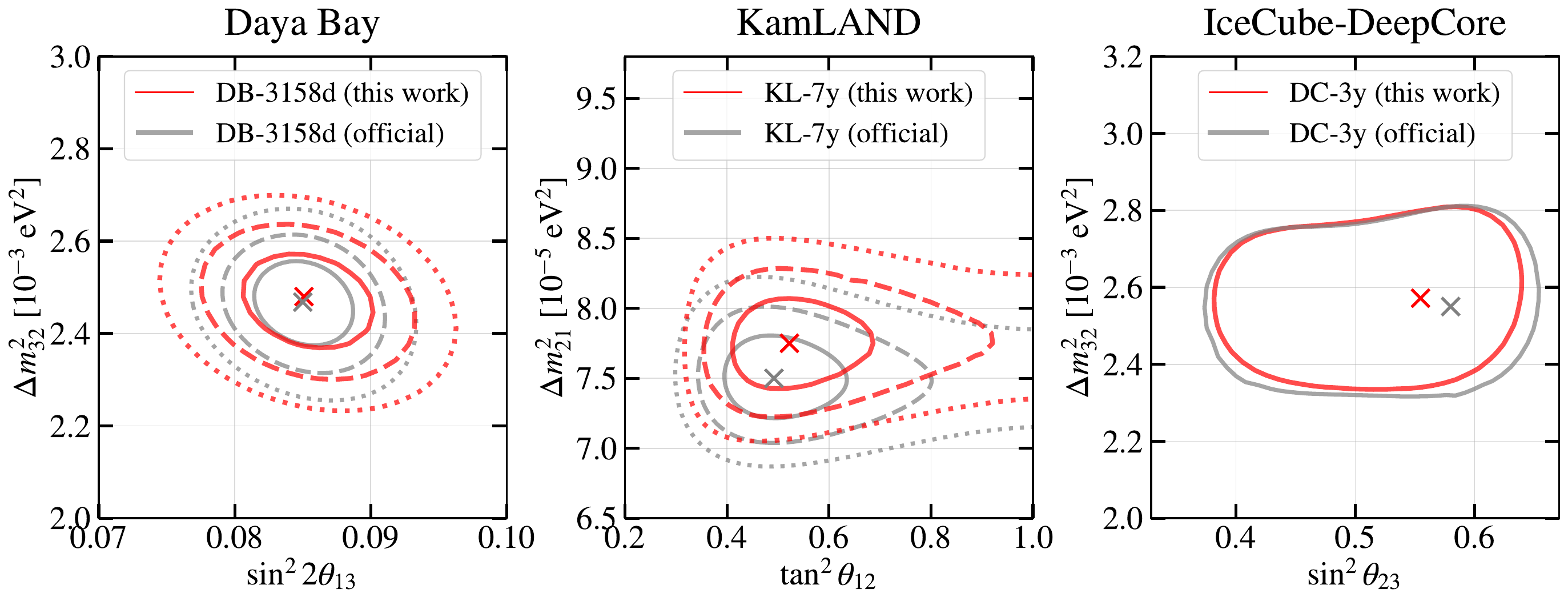}
    \caption{Reproduction of the standard oscillation results from Daya Bay \cite{DayaBay:2022orm}, KamLAND \cite{KamLAND:2010fvi}, and IceCube-DeepCore \cite{IceCube:2019dqi} experiments. For Daya Bay and KamLAND, the solid, dashed, and dotted contours indicate 1$\sigma$, 2$\sigma$, and 3$\sigma$ confidence levels. For IceCube-DeepCore, the contours correspond to the 90\% confidence level. The crosses represent the best fit points from this work (red) or the published experimental result (gray). The official IceCube contours were extracted from \cite{IceCube:2019dqi} (2019 result).}
    \label{fig:three_flavour_reproduction_current}
\end{figure}

\begin{figure}[htb!]
    \centering
    \includegraphics[width=0.7\textwidth]{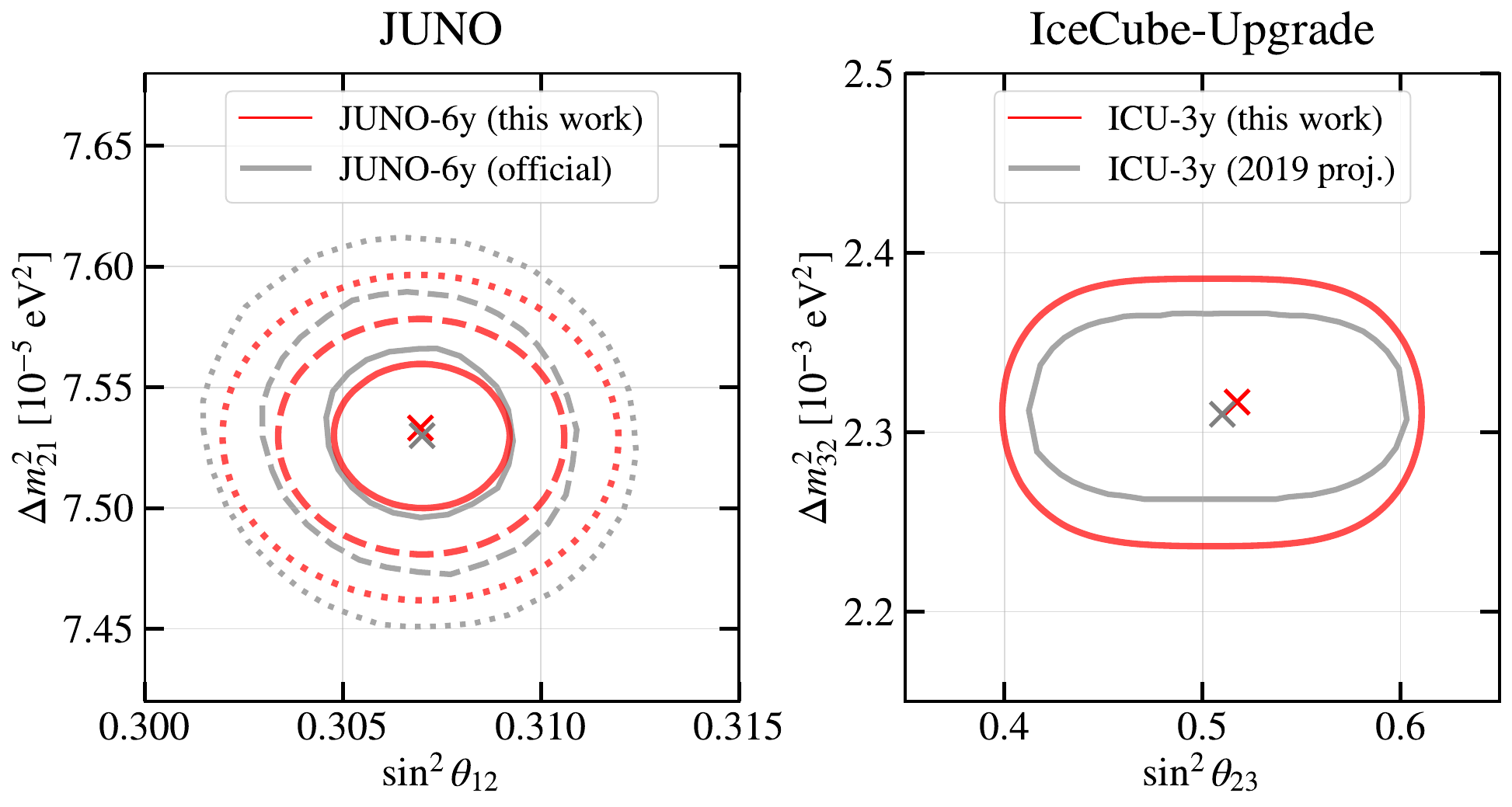}
    \caption{Reproduction of the standard oscillation results from JUNO \cite{JUNO:2022mxj} and IceCube-Upgrade \cite{Ishihara:2019aao} experiments. For JUNO, the solid, dashed, and dotted contours indicate 1$\sigma$, 2$\sigma$, and 3$\sigma$ confidence levels. For IceCube-Upgrade, the contours correspond to the 90\% confidence level. The crosses represent the best fit points from this work (red) or the injected truth (gray).}
    \label{fig:three_flavour_reproduction_future}
\end{figure}

Our results are summarized in \cref{fig:three_flavour_reproduction_current,fig:three_flavour_reproduction_future}.
Generally, we find a good agreement between the official collaboration contours and our reproductions, and note the following:
\begin{itemize}
    \item For Daya Bay, our recovered $2\sigma$ and $3\sigma$ contours are \SIrange[range-phrase=--,range-units=single]{30}{35}{}\% wider than the official result of the Daya Bay collaboration. This level of agreement is expected, as we implemented an approximate covariance matrix with the simplifications detailed in \cref{sec:dayabay,sec:db_covariance}, rather than fully profiling the systematic parameters as done in \cite{DayaBay:2022orm}.
    \item For KamLAND, the best-fit mass splitting $\Delta m_{21}^2$ obtained in our reproduction is 3\% (0.8$\sigma$) higher than the collaboration reported-value, and the best-fit $\tan^{2} \theta_{12}$ is 6\% (0.2$\sigma$) higher than the official KamLAND result. All three confidence levels contours are similarly shifted to higher $\Delta m_{21}^2$ and lower $\tan^2 \theta_{12}$. We attribute this to the difference in the likelihood implementations, namely the binned and time-independent Poisson likelihood used in this study as opposed to the unbinned likelihood incorporating the time-dependent reactor power, which was used in the official KamLAND fit \cite{KamLAND:2008dgz}.
    \item The ($\sin^2 \theta_{23}, \Delta m^2_{32}$) contours derived in this work are in a good agreement with the official IceCube result \cite{IceCube:2019dqi}. We achieve a nearly perfect match of the contour widths in the $\Delta m^2_{32}$ dimension and $\lesssim$20\% discrepancy in the $\sin^2 \theta_{23}$ dimension, with our contours being narrower. This could be connected with the cross section systematic uncertainties being omitted in our analysis and the resulting shift of the best-fit point by $\Delta \sin^2 \theta_{23} = 0.026$.
    
    \item The $2\sigma$ and $3\sigma$ contours obtained in the $(\sin^2 \theta_{12},\,\Delta m_{21}^2)$ space for JUNO are underestimated only at 0.1\%-0.2\% level in our reproduction compared the official JUNO projection \cite{JUNO:2022mxj}. This is expected since we considered only geoneutrinos as a source of background in JUNO, and wider contours would be obtained with more backgrounds and associated systematic uncertainties incorporated into the analysis.
    \item The 90\% confidence level contours derived in this work for the IceCube-Upgrade are \SIrange[range-phrase=--,range-units=single]{13}{18}{}\% wider in the $\sin^2 \theta_{23}$ dimension and \SIrange[range-phrase=--,range-units=single]{40}{50}{}\% wider in the $\Delta m^2_{32}$ dimension, as compared to the official projections from \cite{Ishihara:2019aao}. This is caused by the different form of the test statistic, such that the uncertainty due to limited Monte Carlo statistics is taken into account in this work but not in \cite{Ishihara:2019aao}. We confirmed that the contours match exactly if the Monte Carlo statistics uncertainty term is excluded from our test statistic (see \cref{eq:per_bin_stat_significance}).
\end{itemize}

\section{Global fit}\label{sec:global_fit}

\subsection{Setup}\label{sec:global_fit_setup}

For the primary analysis of this work, we take a Bayesian approach to constrain the individual elements of the leptonic mixing matrix, including their magnitudes and phases. We combine the likelihoods defined in \cref{sec:experiments} for each experiment with the uniform [0, $2\pi$] prior on each of the fitted phases $\phi_{\alpha i}$ and the uniform [0, 1] prior on most of the fitted magnitudes $|N_{\alpha i}|$. The exceptions are $|N_{e1}|$, $|N_{e2}|$, which we constrain to be in the [0.7, 0.9] and [0.4, 0.7] ranges, respectively, using the prior information from the solar oscillation experiments as discussed in \cref{sec:kamland}. The constraints for the mass splittings are $\Delta m^{2}_{21} \in [6.5 \cdot 10^{-5}\,\mathrm{eV^2},\,9.0 \cdot 10^{-5}\,\mathrm{eV^2}$] and $\Delta m^{2}_{31} \in [1.9 \cdot 10^{-3}\,\mathrm{eV^2},\,2.9 \cdot 10^{-3}\,\mathrm{eV^2}$]. Normal neutrino mass ordering is assumed throughout this study\footnote{We tested the possibility of fitting the normal mass ordering model to pseudodata with injected inverted ordering and found virtually identical sensitivities to the mixing matrix elements, normalisations, and closures with the current set of experiments (IceCube-DeepCore, Daya Bay, KamLAND). }. The priors on the IceCube-DeepCore and the IceCube-Upgrade systematic parameters are defined in \cref{sec:atmo_systematics}.

Our fit is performed using the \textsc{UltraNest} \cite{Buchner:2021cql,Buchner:2024ult} package, which utilizes the MLFriends Monte Carlo nested sampling algorithm \cite{Buchner:2014xxx,Buchner:2019xxx}. We couple the likelihood evaluations to the oscillation probability calculations within the \textsc{Neurthino} package \cite{Neurthino:2021xxx}, which we modified to include the physics specific to non-unitarity (see \cref{sec:nonunitary_osc_in_matter,sec:oscillograms_and_normalisation_effects}). The sampling is performed three times: 
\begin{enumerate}[label={(\arabic*)}]
  \item under the \textit{\textbf{agnostic}} assumption (i.e., unrestricted normalisations of the mixing matrix $N$);
  \item under the assumption of $N$ being a \textit{\textbf{submatrix}} of a larger unitary matrix (i.e., imposing that the row and column normalisations are smaller than 1);
  \item under the \textit{\textbf{unitarity}} assumption.
\end{enumerate}
 In the latter case, the mixing matrix is parameterised in terms of mixing angles and the $\delta_{\mathrm{CP}}$ phase, which are most conservatively assigned uniform priors in the $[0, 2\pi]$ range. Thereafter, we impose the same constraints on the elements of the electron row as described earlier for the non-unitary models.

When making sensitivity projections for both the current and the next-generation experiments, we use the unitary mixing matrix $U$ generated with the NuFit 5.2 oscillation parameters \cite{NuFit:2022xxx,Esteban:2020cvm} as the truth. We then constrain the individual matrix elements of $N$ by fitting the non-unitary model to the fake data templates produced with $U$.

\subsection{Current constraints on the non-unitarity parameters}\label{sec:main_results_current}

\subsubsection{Matrix elements, normalisations, and closures}\label{sec:nonunitarity_metrics_results}

\begin{figure}[htb!]
    \centering
    \includegraphics[width=0.95\textwidth]{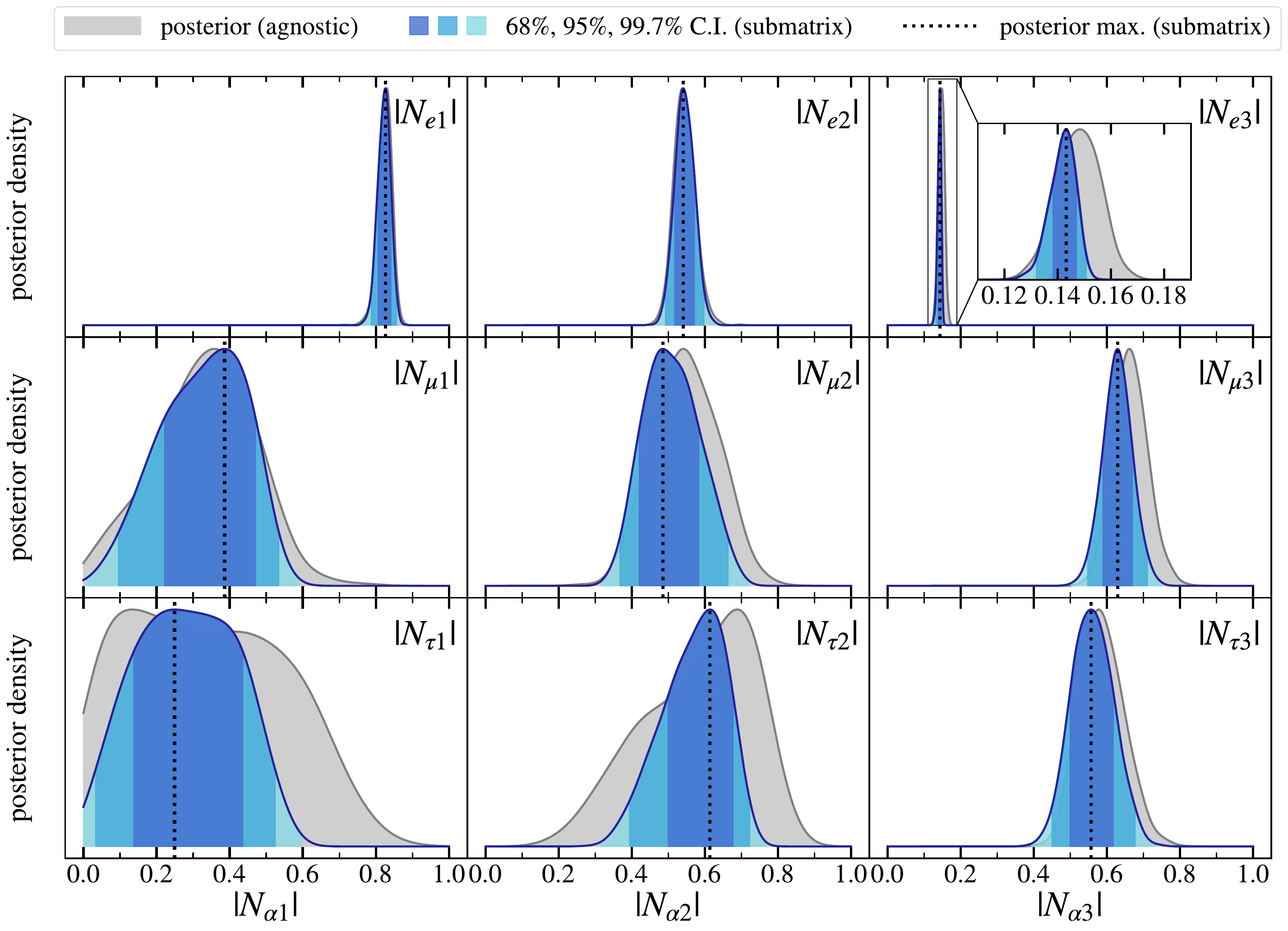}
    \caption{Posterior densities of the non-unitary matrix element magnitudes, $|N_{\alpha i}|$, as obtained with the current data from IceCube-DeepCore, Daya Bay, and KamLAND. All other fitted physics and nuisance parameters are marginalised over. The full posterior for the \textit{\textbf{agnostic}} case is shown in gray. For the \textit{\textbf{submatrix}} case, the three shaded blue regions correspond to 68\%, 95\%, and 99.7\% credible intervals (C.I.), from darker to lighter.}
    \label{fig:individual_elements_current}
\end{figure}

In \cref{fig:individual_elements_current}, we show the posterior densities of each of the matrix element magnitudes $|N_{\alpha i}|$ obtained with the available public data from IceCube-DeepCore, Daya Bay, and KamLAND. These results are initially provided for both the agnostic model (unrestricted row and column normalisations) and the submatrix model ($N_{\alpha},\,\,N_i < 1$), while in later sections we focus on the physically motivated submatrix case.

Our first observation is that the electron row elements are significantly better constrained than the muon and the tau row elements. In particular, the few-percent precision of the measured $|N_{e3}|$ is driven by the extremely high statistics of the $\bar{\nu}_{e}$ neutrino sample detected by Daya Bay, while the constraints on $|N_{e1}|$ and $|N_{e2}|$ are the result of combining the Daya Bay data with that of KamLAND and the prior knowledge from the solar neutrino experiments about the octant of $\theta_{12}$. The primary advantage of including the IceCube-DeepCore data into the global fit lies in the measurement $|N_{\mu 3}|$ and $|N_{\tau 3}|$ elements, which we find to be constrained with $\sim$\SIrange[range-phrase=--,range-units=single]{30}{40}\% precision given the three-year dataset. Thanks to the constraints on the overall row normalizations (discussed further in this section), the $|N_{\mu\{1,2\}}|$ and the $|N_{\tau\{1,2\}}|$ elements could also be measured, albeit with large uncertainties (60$\text{-}$100\%, depending on the model) and strong degeneracies between the elements of the first and the second column in each row.

\begin{figure}[htb!]
    \centering
    \includegraphics[width=0.75\textwidth]{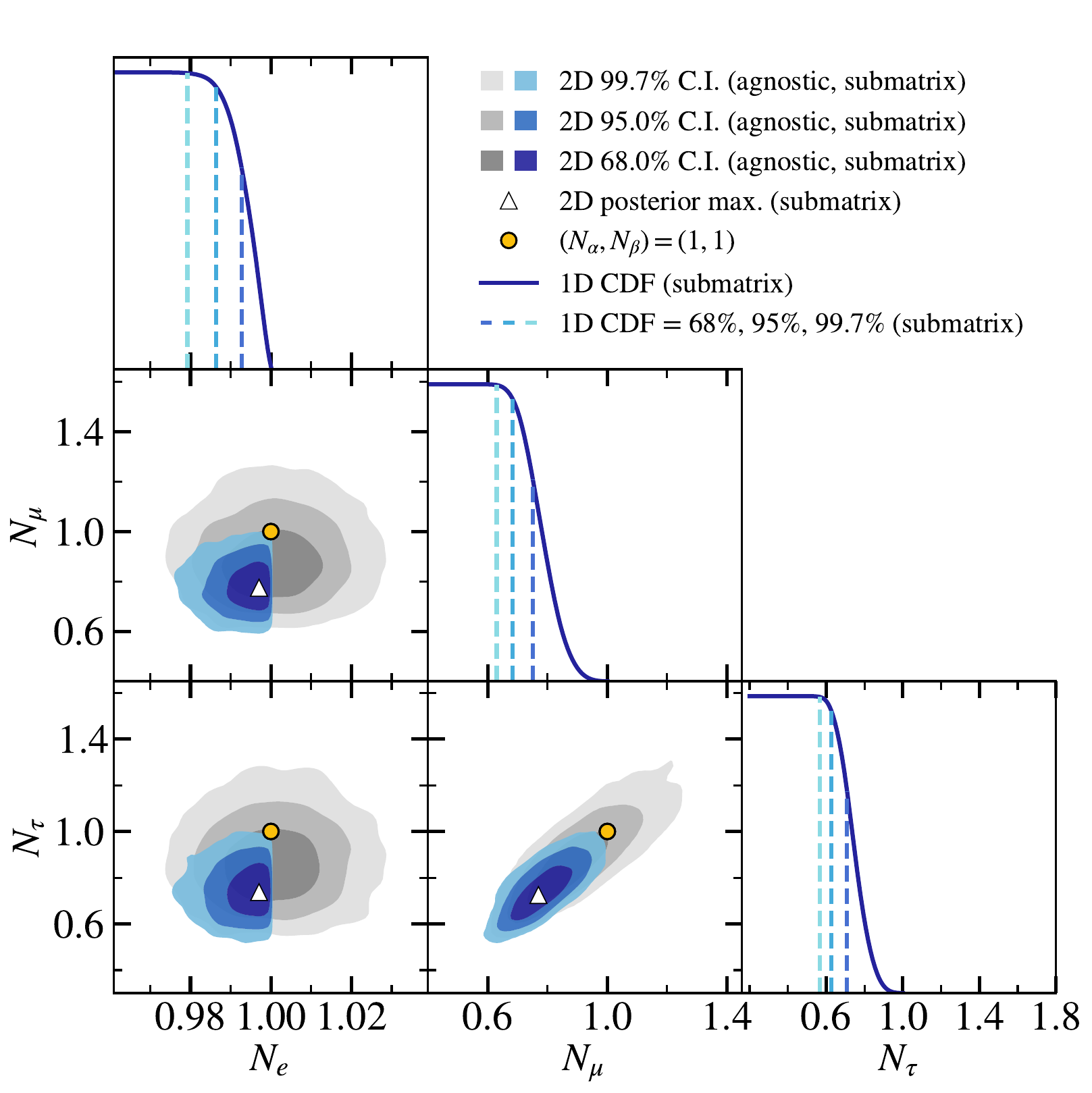}
    \caption{Off-diagonal elements: 2D posteriors of the matrix row normalisations $(N_{\alpha}, N_{\beta})$  as measured under the \textit{\textbf{agnostic}} assumption (gray) and the \textit{\textbf{submatrix}} assumption (blue). Diagonal elements: 1D cumulative distribution functions (CDFs) of the row normalisations $N_{\alpha}$ as measured under the \textit{\textbf{submatrix}} assumption.}
    \label{fig:row_norms_current}
\end{figure}

\begin{figure}[h!]
    \centering
    \includegraphics[width=0.75\textwidth]{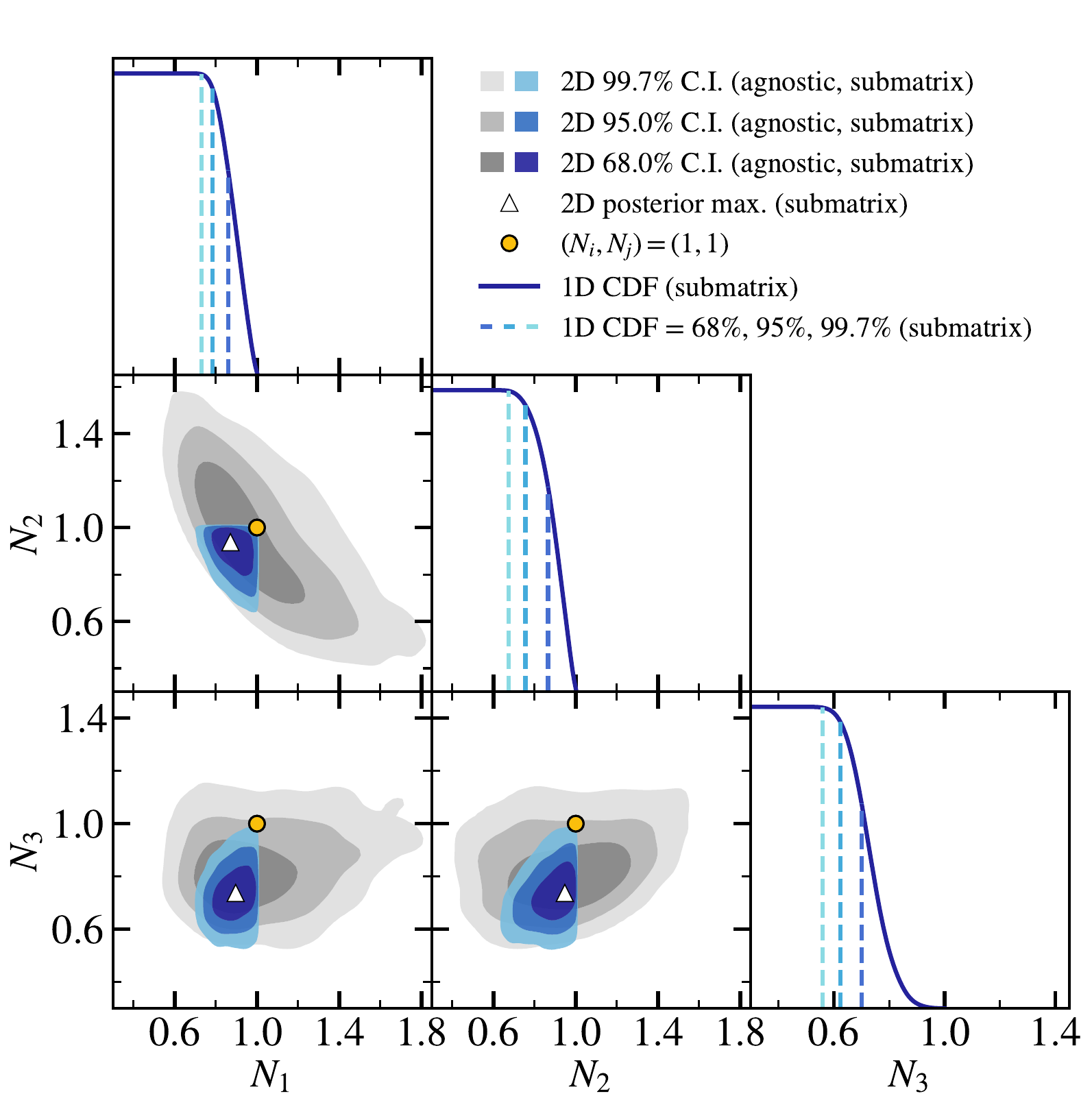}
    \caption{Same as \cref{fig:row_norms_current}, but applied to the matrix columns $(i, j)$ instead of the matrix rows $(\alpha, \beta)$.}
    \label{fig:col_norms_current}
\end{figure}

From the posterior densities of $|N_{\alpha i}|$, we construct the posteriors of the row normalisations, $N_{\alpha}$, and the column normalisations, $N_i$. The results are shown in \cref{fig:row_norms_current,fig:col_norms_current}, respectively. The one-sided 68\%, 95\%, and 99.7\% credible intervals derived from the normalisation posteriors in the submatrix case are additionally reported in \cref{tab:normalisation_credible_intervals}. These results can be compared to the much more stringent constraints not involving neutrino oscillations, e.g. those obtained via the searches for flavour-violating decays of charged leptons, electroweak universality tests, and similar \cite{Antusch:2014woa,Fernandez-Martinez:2016lgt,deGouvea:2015euy,Blennow:2016jkn,Blennow:2023mqx}. Such constraints are typically presented in the ``$\alpha$'' parameterisation of the non-unitary mixing matrix, that is, assuming $N = (\mathbb{I} - \hat{\alpha}) U$. We use the constraints on $\hat{\alpha}$ from \cite{Blennow:2023mqx} to derive the corresponding constraints on the row and column normalisations, which we add to \cref{tab:normalisation_credible_intervals} as a reference. 

\begin{table}[!htbp]
\centering
\renewcommand{\arraystretch}{1.35}
\begin{tabular}{lcccc}
\hline
           & 68\% C.I.          & 95\% C.I.          & 99.7\% C.I.   & EW (95\% C.L.; \cite{Blennow:2023mqx})    \\ \hhline{=====}
$N_e$      & 0.993 & 0.987 & 0.979 & 0.9972 \\ \hline
$N_{\mu}$  & 0.75   & 0.68   & 0.63  & 0.9997 \\ \hline
$N_{\tau}$ & 0.71   & 0.63   & 0.57  &  0.998 \\ \hline
$N_1$      & 0.86   & 0.79   & 0.73  & 0.996 \\ \hline
$N_2$      & 0.87   & 0.76   & 0.68  & 0.996 \\ \hline
$N_3$      & 0.70   & 0.62   & 0.56  &  0.996 \\ \hline
\end{tabular}
\caption{First three columns: Lower bounds of the one-sided credible intervals (C.I.) for the 1D posteriors of the row normalisations $N_{\alpha}$ and the column normalisations $N_{i}$ of the neutrino mixing matrix $N$, as measured in this study under the \textit{\textbf{submatrix}} assumption. Last column: lower bounds on the unitarity constraints derived from the global fits to the precision electroweak and flavour observables (shortly labeled as ``EW'') \cite{Blennow:2023mqx}. Note that the EW constraints on the column normalisations $N_i$ are identical for all columns, following the procedure from \cite{Antusch:2006vwa}.}
\label{tab:normalisation_credible_intervals}
\end{table}

From the present study, we find that the electron row normalisation is well constrained by the current data, such that the 99.7\% credible interval extends only 2\% below the unitary expectation of $N_e = 1$. We remind the reader that this constraint comes predominantly from the Daya Bay data and is the result of our specific choice of the analysis where the input IBD cross section is based on the Standard Model calculation from \cite{Vogel:1999zy}. This necessitates the correction of the Daya Bay oscillation probabilities by one factor of $N_e$, as per \cref{eq:effective_osc_prob_dayabay}, which would have cancelled out in an analysis relying on near-to-far detector event ratios but does not cancel out in a direct calculation of the expected event rates with a SM cross section. Our constraint on $N_e$ therefore represents the statistical power of Daya Bay to constrain the $\bar{\nu}_e$ spectrum normalisation affected by the non-unitarity effects on the cross section. If no extra factors of $N_e$ were modifying the effective oscillation probability, the non-unitarity in the electron row would only be observed by KamLAND due to the presence of non-standard matter effects, but would not be observed in Daya Bay. 

Further, the derived normalisations of the muon and the tau rows have much wider contours than those of the electron row, covering $\sim$$70\%$ ranges about ($\sim$$30\text{-}40\%$ ranges below) the unitary expectation in the agnostic (submatrix) case.  The respective posterior maxima are situated at $(N_{\mu},\,N_{\tau}) = (0.77, 0.73)$ in the submatrix case \footnote{For reference, the value of the $\nu_{\tau}$ normalisation reported in the official IceCube-DeepCore tau neutrino appearance result was $0.73^{+0.30}_{-0.24}$ ($0.57^{+0.36}_{-0.30}$) and obtained through a simple scaling of the expected number of the $\nu_{\tau}$-CC (CC+NC) events \cite{IceCube:2019dqi}. }. We note a strong positive correlation between the $N_{\mu}$ and $N_{\tau}$ posterior densities. In \cref{sec:atmo_sys_impact}, we reveal that these correlations are driven by the atmospheric neutrino systematic uncertainties and would not have existed if the atmospheric neutrino flux was perfectly constrained. On the contrary, the negative correlation between the $N_1$ and $N_2$ column normalisations is the result of the degeneracies between $|N_{\alpha 1}|$ and $|N_{\alpha 2}|$ ($\alpha \in [\mu, \tau]$), which arise due to the fact that the atmospheric oscillation probabilities only depend on the sum of squares of these elements \cite{Ellis:2020hus}. 

\begin{figure}[h!]
    \centering
    \includegraphics[width=0.75\textwidth]{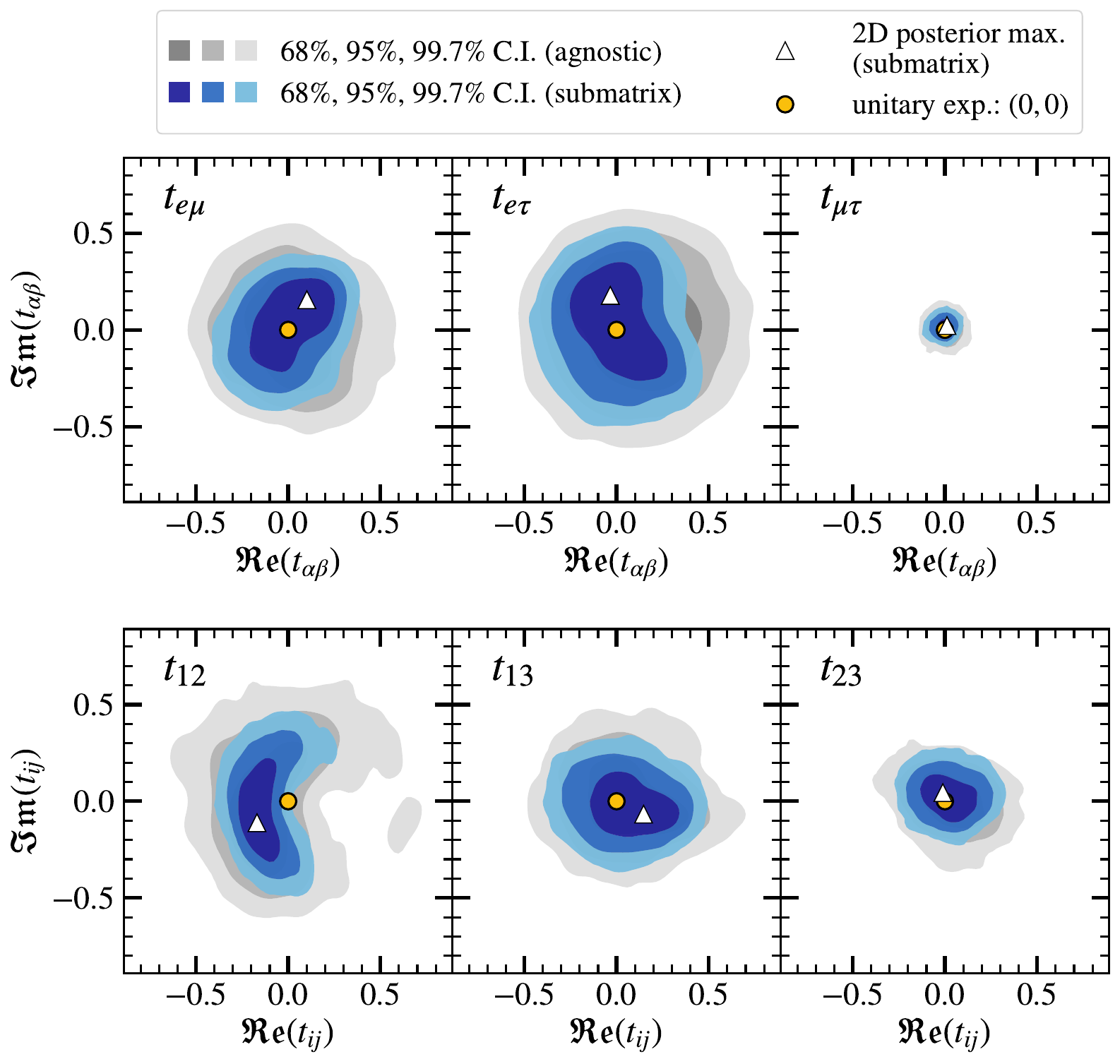}
    \caption{\textit{Top}: posteriors of the row closures $t_{\alpha\beta}$ as obtained with the current data from IceCube-DeepCore, Daya Bay, and KamLAND. The results are displayed in the 2D complex planes for the two considered models: \textit{\textbf{agnostic}} (gray) \textit{vs.} \textit{\textbf{submatrix}} (blue). \textit{Bottom}: same as top, but applied to the column closures $t_{ij}$.}
    \label{fig:closures_current}
\end{figure}

Finally, we make use of the fitted complex phases $\phi_{\alpha i}$ in combination with the matrix element magnitudes $|N_{\alpha i}|$ to derive the constraints on the row closures $t_{\alpha\beta}$ and the column closures $t_{ij}$, as defined in \cref{sec:oscillograms_and_normalisation_effects}. The 2D posteriors in the complex ($\mathfrak{Re}$, $\mathfrak{Im}$) planes for each of the closures and both of the considered models are given in \cref{fig:closures_current}. Most of the closure posteriors enclose the unitary expectation of $\mathfrak{Re}(t_{\ldots}) = \mathfrak{Im}(t_{\ldots}) = 0$ within the 68\% credible interval, except for $t_{12}$, which contains the unitary expectation within the 95\% (99.7\%) credible interval in the agnostic (submatrix) case. Thus, our measurements of the row and column closures demonstrate consistency with unitarity. The best constraints are obtained for $t_{\mu\tau}$, such that the 99.7\% credible interval for $|t_{\mu\tau}|$ is contained within the radius of $0.1\text{-}0.2$ from the unitary expectation.

\subsubsection{Comparison with the model assuming unitarity}\label{sec:unitarity_comparison_bayes_factors}

In \cref{sec:main_results_current}, we showed that the data appears to prefer smaller-than-unity normalisations $N_{\mu}$, $N_{\tau}$, and $N_3$ in both agnostic and submatrix scenarios. We now seek to contrast these non-unitary models with the one that assumes unitarity, i.e., the standard three-flavour neutrino mixing through the PMNS matrix as defined in \cref{eq:unitary_matrix_mixing_angles}. For that purpose, we run an additional Bayesian fit for the unitary scenario (as described in \Cref{sec:global_fit_setup}) and use \textsc{Ultranest} to extract the Bayesian evidence $\mathcal{Z}$ for all three models. The evidence is computed as
\begin{equation}
    \mathcal{Z} = \int \mathcal{L}(\mathcal{N}_{\mathrm{obs}}| \bar{\lambda}_{\mathrm{osc}}, \bar{\lambda}_{\mathrm{syst}}) \pi(\bar{\lambda}_{\mathrm{osc}}, \bar{\lambda}_{\mathrm{syst}}) \mathrm{d}\bar{\lambda}_{\mathrm{osc}}\mathrm{d}\bar{\lambda}_{\mathrm{syst}},
\label{eq:evidence}
\end{equation}
where $\mathcal{L}(\mathcal{N}_{\mathrm{obs}}| \bar{\lambda}_{\mathrm{osc}}, \bar{\lambda}_{\mathrm{syst}})$ is the product of the per-experiment likelihoods combined in the global fit, $\pi(\bar{\lambda}_{\mathrm{osc}}, \bar{\lambda}_{\mathrm{syst}})$ are the priors for the oscillation (systematic) parameters $\bar{\lambda}_{\mathrm{osc}}$ ($\bar{\lambda}_{\mathrm{syst}}$), and $\mathcal{N}_{\mathrm{obs}}$ are the observed data. Having obtained the evidence values $\mathcal{Z}_{\mathrm{agn.}}$, $\mathcal{Z}_{\mathrm{submat.}}$, and $\mathcal{Z}_{\mathrm{SM}}$ for the agnostic, submatrix, and the unitary (``Standard Model'') scenarios, we compute the logarithms of the Bayes factors as
\begin{equation}
    \ln \mathcal{B}_{\mathrm{agn.}} = \ln \mathcal{Z}_{\mathrm{SM}} - \ln \mathcal{Z}_{\mathrm{agn.}};\qquad \ln \mathcal{B}_{\mathrm{submat.}} = \ln \mathcal{Z}_{\mathrm{SM}} - \ln \mathcal{Z}_{\mathrm{submat.}},
\end{equation}
such that negative (positive) values of $\ln \mathcal{B}$ indicate a preference for the non-unitary (unitary) model. We obtain $\ln \mathcal{Z}_{\mathrm{agn.}} = -125.8$, $\ln \mathcal{Z}_{\mathrm{submat.}} = -125.6$, and $\ln \mathcal{Z}_{\mathrm{SM}} = -111.8$, resulting in $\ln \mathcal{B}_{\mathrm{agn.}} = 14.0$ and $\ln \mathcal{B}_{\mathrm{submat.}} = 13.8$. According to the Jeffreys' scale \cite{Jeffreys:1939XXX,Trotta:2008qt}, this implies a strong preference for the unitary model over the non-unitary ones. This finding is in contrast with the fact that the agnostic and the submatrix model fits prefer off-unitary normalisations for certain rows and columns, as indicated above. Given that the non-unitary models introduce 9 extra oscillation parameters compared to the unitary case, the resulting Bayes factors can simply reflect the relative complexity of the models -- i.e., the non-unitary models being unnecessarily complex. To get further insight into whether the best-fit unitary and the best-fit non-unitary models are individually compatible with the data, we run posterior predictive checks for each model as described in \cref{sec:goodness_of_fit}.

\subsubsection{Posterior predictive checks}\label{sec:goodness_of_fit}

\begin{figure}[h!]
    \centering
    \includegraphics[width=\textwidth]{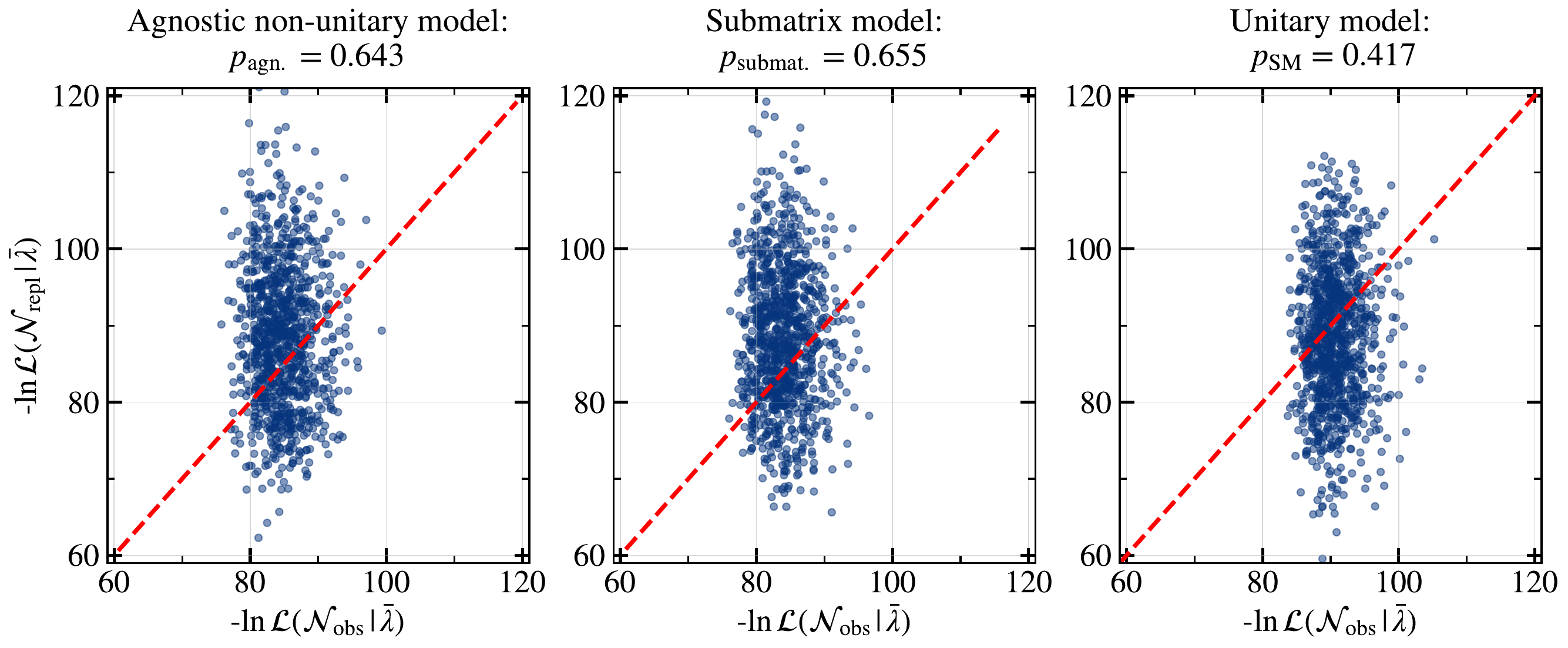}
    \caption{The results of the posterior predictive checks via realised discrepancies for the agnostic non-unitary model (left), the submatrix model (middle), and the unitary model (right). In each of the scatter plots, the fraction of the points lying above the 1:1 line corresponds to the $p$-value for a given model. More details on the method are provided in text and Ref.~\cite{Gelman:1996xxx}.}
    \label{fig:ppc_pvalues}
\end{figure}

To quantify the goodness of fit for both the unitary and the non-unitary models tested in this study, we use the method of posterior predictive checks via realised discrepancies, as devised in \cite{Gelman:1996xxx}. The purpose of the method is to assess whether the obtained posterior distributions of the fitted parameters can yield observations that ``cover'' the original data. A successful outcome of such a test would be that the data does not deviate from the expectation under a given set of parameters more than the majority of the statistical realisations of the same expectation. This test is repeated for every posterior sample of $\bar{\lambda} = \{\bar{\lambda}_{\mathrm{osc}}, \bar{\lambda}_{\mathrm{syst}}\}$, which can be used to generate an expectation template $\mathcal{N}_{\mathrm{exp}}$ for each experiment and further statistically fluctuated (e.g. by applying Poisson fluctuations) to obtain a mock ``replica'' of the data $\mathcal{N}_{\mathrm{repl}}$. Then, given the current posterior sample of $\bar{\lambda}$, the log-likelihoods $\ln \mathcal{L}(\mathcal{N}_{\mathrm{data}}\,|\,\bar{\lambda})$ and $\ln \mathcal{L}(\mathcal{N}_{\mathrm{repl}}\,|\,\bar{\lambda})$ can be computed for the original data and the mock replica. One can eventually claim that the posteriors result in replicas that cover the data well if $-\ln \mathcal{L}(\mathcal{N}_{\mathrm{repl}}) > -\ln \mathcal{L}(\mathcal{N}_{\mathrm{data}})$ for a sufficiently large fraction of $\bar{\lambda}$ sampled from the posterior \cite{Gelman:1996xxx}. This fraction is then defined as the $p$-value describing the compatibility between the tested model with its posteriors and the data. In \cref{fig:ppc_pvalues}, we plot the $-\ln \mathcal{L}(\mathcal{N}_{\mathrm{repl}})$ values against the $-\ln \mathcal{L}(\mathcal{N}_{\mathrm{data})}$ values obtained for 1000 equally weighted posterior samples pertaining to each of the two models. The $p$-values are found as the fractions of points above the 1:1 lines, which results in $p_{\mathrm{agn.}} \simeq 0.64$, $p_{\mathrm{submat.}} \simeq 0.66$, and $p_{\mathrm{SM}} \simeq 0.42$. 

\noindent We deem all of these $p$-values to be acceptable and conclude that the data is well described by both non-unitary and unitary models. This further strengthens our earlier argument in \cref{sec:unitarity_comparison_bayes_factors} that the strong preference for the unitary model according to the Bayes factor is likely a mere consequence of the non-unitary models penalised for their complexity.

\subsection{Impact of atmospheric neutrino systematics}\label{sec:atmo_sys_impact}

\begin{figure}[h!]
    \centering
    \includegraphics[width=0.7\textwidth]{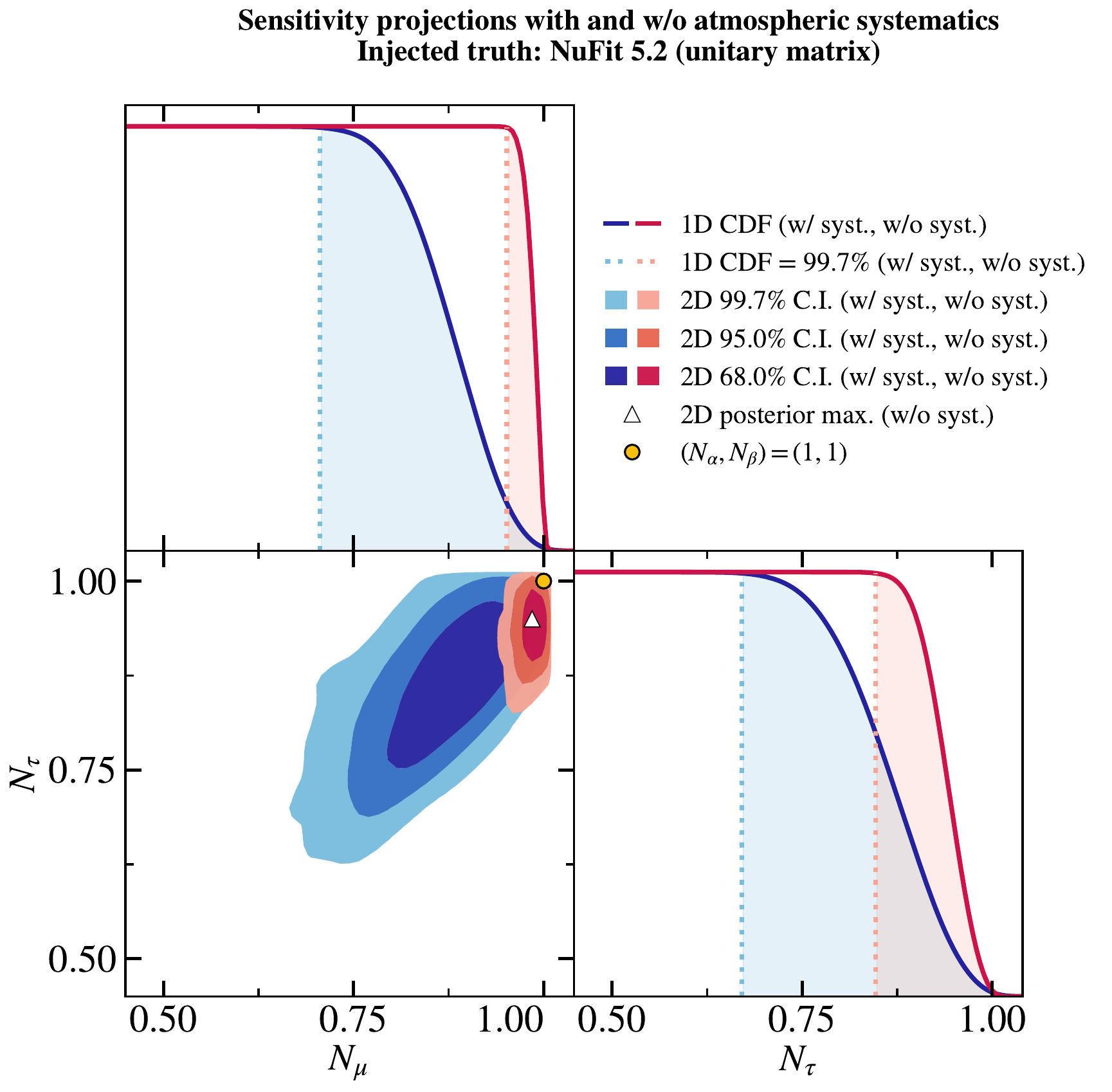}
    \caption{The projected posterior densities of the row normalisations $N_{\mu}$ and $N_{\tau}$ with and without atmospheric systematic uncertainties (shown as blue and red contours, respectively), as obtained under the \textbf{\textit{submatrix}} assumption.}
    \label{fig:atmo_sys_impact}
\end{figure}

To test the impact of the atmospheric neutrino systematic uncertainties on the non-unitarity metrics evaluated in this study, we perform two additional Bayesian fits. In both of the fits, we replace the real data from the considered experiments with the pseudodata templates generated under the assumption of unitarity\footnote{Fitting pseudodata rather than real data is necessary for this test since the data would not be well described without the systematic uncertainty parameters. This would result in a bad goodness of fit and complicate the interpretation of the test outcomes.} and with the injected oscillation parameters fixed at the NuFit 5.2 values  \cite{Esteban:2020cvm,NuFit:2022xxx}. In one of the fits, we further exclude the IceCube-DeepCore systematic uncertainties from consideration by fixing their values at the nominal expectations \cite{IceCube:2019dqi}. The results of these two fits are shown in \cref{fig:atmo_sys_impact}.

We find that if all of the IceCube-DeepCore systematic parameters were known perfectly, the constraints on $N_{\mu}$ and $N_{\tau}$ normalisations individually would be much stronger. In particular, in the submatrix case the width of the one-sided 99.7\% credible interval of $N_{\mu}$ would shrink by a factor of 6 (from 0.3 to 0.05), and that of $N_{\tau}$ -- by a factor of 2 (from 0.3 to 0.15). Furthermore, the correlation between $N_{\mu}$ and $N_{\tau}$ would be broken, resulting in constraints similar to those obtained in \cite{Ellis:2020hus}.

To better understand which of the nuisance parameters induce the ($N_{\mu}$, $N_{\tau}$) correlations and lead to the wide one-dimensional $N_{\mu}$ and $N_{\tau}$ posteriors, we compare the nuisance parameter posterior densities between the non-unitary and the unitary model fits to the current data (see \cref{fig:current_sys_params_fit_results}). The difference in the posterior distributions obtained with the two models for each parameter allows us deduce which of the IceCube-DeepCore parameters are degenerate with non-unitarity. In particular, the shift in the means of the distributions between the unitary and the non-unitary model fits suggests that the combination of the parameters in question is degenerate with the shift of the $N_{\mu}$ and $N_{\tau}$ posterior maxima to the off-unitary values. On the contrary, the change in the distribution widths of the individual systematic parameters implies that these parameters are responsible for the spread of the $N_{\mu}$ and $N_{\tau}$ distributions (in the unitary case, the spread is equal to 0, as $N_{\mu} = N_{\tau} = 1$). The two parameters with the largest change in the distribution width are the overall normalisation, whose 99.7\% credible interval shrinks by a factor 3.4 when unitarity is enforced, and the spectral index of the neutrino flux, which shrinks by 21\%. These are followed by the overall efficiency of the optical modules (10\% change in the distribution width) and the relative normalisation of the NC events (9\% change). By directly plotting the posteriors of $N_{\mu}$, $N_{\tau}$ against these four systematic parameters in \cref{fig:nmu_ntau_correlations_with_sys_params} and calculating the respective Pearson correlation coefficients, we confirm that the strongest correlations are indeed observed between $N_{\mu}$, $N_{\tau}$ and the two leading atmospheric neutrino flux parameters (normalisation and spectral index). We therefore conclude that tightening the priors on these parameters would help place stronger constraints on the $N_{\mu}$ and $N_{\tau}$ normalisations.

\subsection{Future projections}\label{sec:future_projections}

\begin{figure}[htb!]
    \centering
    \includegraphics[width=0.95\textwidth]{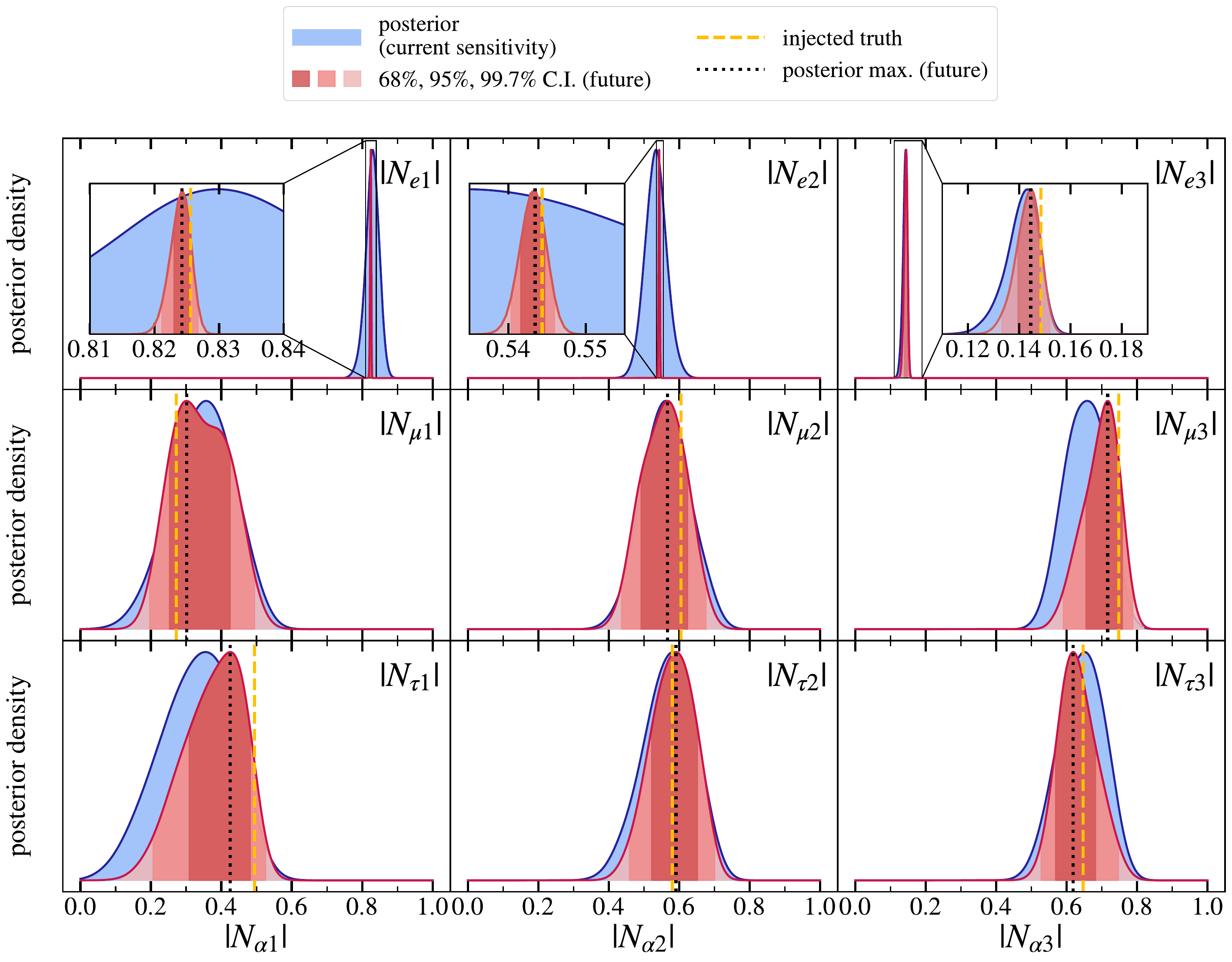}
    \caption{Same as \cref{fig:individual_elements_current}, but applied to the future projections of the matrix element magnitudes using the combined IceCube-Upgrade, JUNO, and Daya Bay data fitted with the \textit{\textbf{submatrix}} model. The injected truth is the unitary matrix generated with NuFit 5.2 oscillation parameters \cite{Esteban:2020cvm,NuFit:2022xxx}.}
    \label{fig:individual_elements_future}
\end{figure}

\begin{figure}[htb!]
    \centering
    \includegraphics[width=0.75\textwidth]{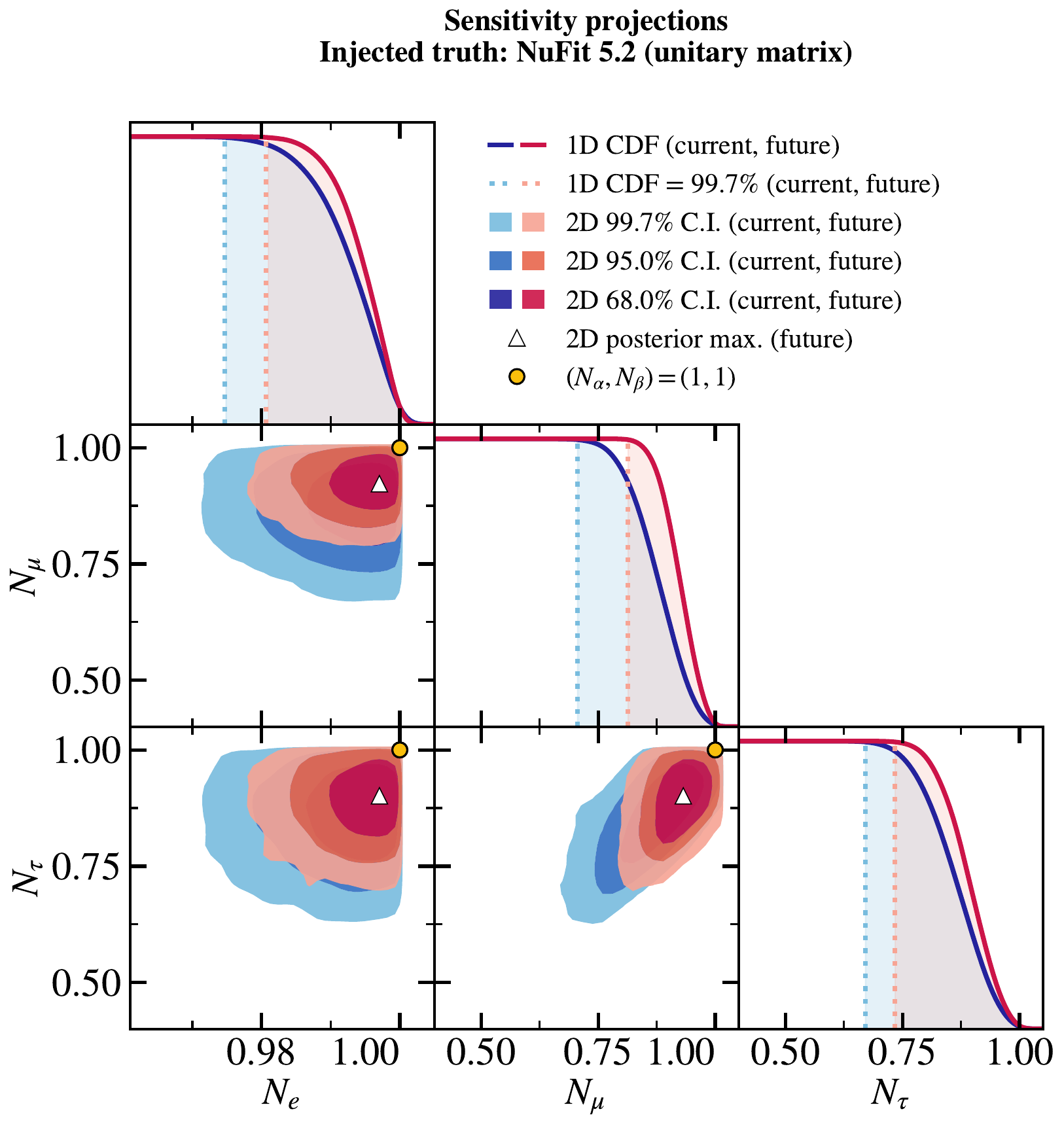}
    \caption{Same as \cref{fig:row_norms_current}, but applied to the future projections of the matrix row normalisations using the combined IceCube-Upgrade, JUNO, and Daya Bay data fitted with the \textit{\textbf{submatrix}} model. The injected truth is the unitary matrix generated with NuFit 5.2 oscillation parameters \cite{Esteban:2020cvm,NuFit:2022xxx}.}
    \label{fig:row_norms_future}
\end{figure}

With the upcoming IceCube-Upgrade and JUNO detectors, the constraints on both the individual mixing matrix elements and the non-unitarity metrics will be improved, as shown in \cref{fig:individual_elements_future,fig:row_norms_future} for the submatrix case.
A particular enhancement is seen in the sensitivity to the $|N_{e1}|$ and $|N_{e2}|$ elements, with more than an order of magnitude decrease in the width of the respective 99.7\% credible intervals. The width of the projected $|N_{e3}|$ element posterior does not change significantly with the inclusion of JUNO, as it is constrained predominantly by the high-statistics IBD selection of Daya Bay. Similarly, the $N_e$ normalisation constraint is mainly driven by the Daya Bay data as described in \cref{sec:main_results_current}, and the large quoted improvement in the individual sensitivities to $|N_{e1}|$ and $|N_{e2}|$ leads only to a $\sim$25\% tighter $N_e$ constraint due to the anticorrelations between these elements. The constraints on the muon and the tau row normalisations are expected to improve by $\sim$40\% and 20\%, respectively, when replacing the three years of the current IceCube-DeepCore data with three years of the future IceCube-Upgrade data. A similar degree of improvement is projected for the column normalisations and the closures, as seen from \cref{fig:col_norms_future,fig:closures_future}. 

\section{Discussion and outlook}\label{sec:discussion}

The purpose of this study was to test the unitarity of the neutrino mixing matrix by measuring its individual elements through neutrino oscillation data. We continued similar recent efforts in the literature \cite{Parke:2015goa,Ellis:2020hus,Denton:2021mso}, which combined multiple experiments into global fit analyses to access all of the elements of the mixing matrix. However, rather than building on these previous studies ``horizontally,'' i.e., by expanding the scope of the considered experimental datasets, this work advanced the approach to the non-unitarity analysis ``vertically'' for a selected subset of experiments (reactor and atmospheric). This involved performing full fits of the non-unitary mixing model to the available public datasets and including the experimental nuisance parameters into the fits.

Our analysis provides constraints on the electron row normalisation at the level of a few percent and the tau row normalisation at the level of approximately 30-40\%, results that are comparable with the state-of-the-art global fits to neutrino oscillation data \cite{Parke:2015goa,Ellis:2020hus,Denton:2021mso}. We find that the data considered here is well-described by both unitary and non-unitary mixing models, with the respective posterior predictive $p$-values of $\sim$65\% and 42\%. The computed Bayes factor shows a preference for the unitary model by 14 units in log-evidence, indicating that the non-unitary models may be too complex for the current data.

This work is the first to comprehensively analyse atmospheric neutrino data with systematic uncertainties in a global fit for non-unitarity. An important outcome of this effort is the revealed degeneracy between the non-unitarity metrics (namely the muon and the tau row normalisations) and atmospheric neutrino flux systematic uncertainties. The overall flux normalisation and spectral index uncertainty couple these normalisations, making their constraints positively correlated. This correlation, not seen in previous studies, highlights the importance of treating nuisance parameters carefully in new physics searches to avoid overly optimistic or otherwise misleading results. It also motivates the need to better constrain systematic uncertainties, potentially by incorporating the data-driven atmospheric neutrino flux model \cite{Yanez:2023lsy,Fedynitch:2022vty}.

As the row and column normalisations that were found to be in a mild tension with unitarity were constrained predominantly by the three-year IceCube-DeepCore dataset, it remains to be seen whether these results are simply a statistical feature of the latter through future analyses with more atmospheric neutrino data. This could include, for example, the 9-year IceCube-DeepCore dataset \cite{IceCube:2024xjj} or the data from the upcoming IceCube-Upgrade \cite{Ishihara:2019aao,IceCube:2023ins}, as well as other atmospheric neutrino experiments such as KM3NeT or Super-Kamiokande. In the context of future projections, we have shown that the addition of atmospheric neutrino data with a lower ($\mathcal{O}$(\SI{1}{GeV})) energy reach and a $\sim$2 times higher energy resolution -- both possible with the IceCube-Upgrade compared to the IceCube-DeepCore -- will help tighten the constraints on $N_{\mu}$, $N_{\tau}$, and all of the column normalisations by $20\texttt{-}40\%$. Although these projections could be further improved by incorporating recent advancements in the IceCube-Upgrade detector simulation and the low-energy cross section systematic uncertainties, they already provide an estimate of the IceCube-Upgrade capabilities to constrain non-unitarity and set the foundation for future studies in this direction. 
In the electron sector, we found that the main advantage of including the data from the upcoming JUNO experiment will be in tightening the constraints on $|N_{e1}|$ and $|N_{e2}|$ by more than an order of magnitude. The constraint on the electron row normalisation will be improved by nearly 25\% but remain to be dominated by Daya Bay data even in the JUNO era. Similarly to the IceCube-Upgrade, the exact quantification of the sensitivity enhancement expected from JUNO is subject to a more careful treatment of the detector response and the sources of background, which will be possible to understand and model more accurately when the experiment becomes operational.

Going forward, our vision is to expand the non-unitarity analysis both horizontally and vertically, i.e., by adding the solar and the long-baseline experiments into the global fit while implementing the relevant systematic uncertainties for these experiments. This undertaking will be the more successful the more public information is provided by the experimental collaborations, and any other searches for new physics through global fits will similarly benefit from the availability of open data and analysis prescriptions.

\acknowledgments

This work has been supported by the Deutsche
Forschungsgemeinschaft (DFG, German Research Foundation) under the Sonderforschungsbereich (Collaborative Research Center) SFB1258 ‘Neutrinos and Dark Matter in Astro- and Particle Physics’. The authors, T.K. and D.J.K., are supported by the Carlsberg Foundation (project no. 117238) and acknowledge the computational resources and assistance provided by the SCIENCE High
Performance Computing Center at the University of Copenhagen. The authors additionally thank Leonardo Jos\'{e} Ferreira Leite and Kevin A. Urqu\'{i}a-Calder\'{o}n for valuable feedback on the manuscript, Thomas Stuttard for assistance with the IceCube-Upgrade public data release, as well as Johannes Buchner for helpful suggestions on the application of the nested sampling method to this study.

\clearpage

\appendix
\counterwithin{figure}{section}

\section{Non-unitary neutrino oscillations in vacuum}

\begin{figure}[htb!]
    \centering
    \includegraphics[width=0.95\textwidth]{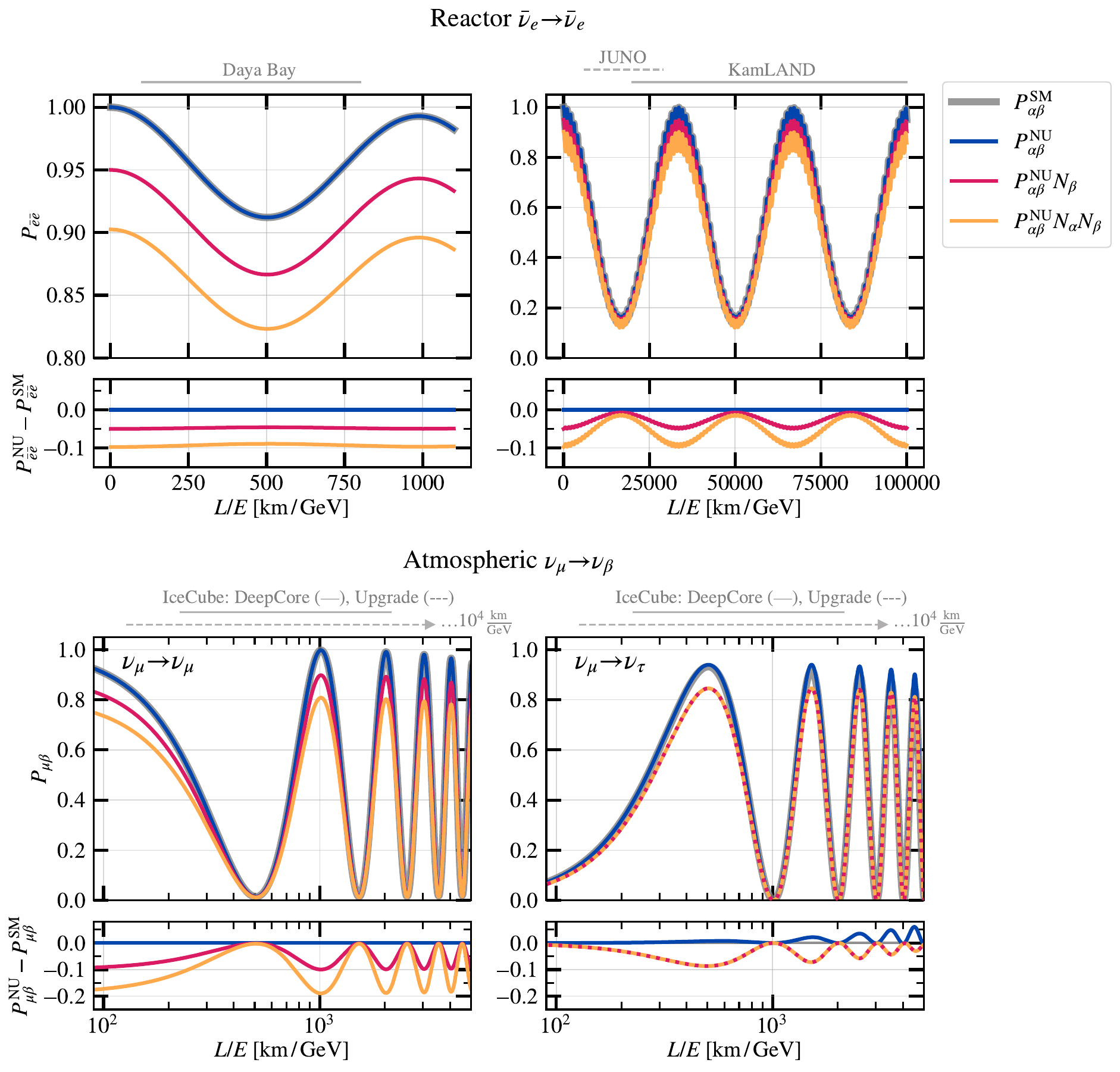}
    \caption{Comparison of the unitary (Standard Model, ``SM'') and the non-unitary (``NU'') oscillation probabilities for the case of $\bar{\nu}_e$ disappearance (top), $\nu_{\mu}$ disappearance (bottom left), and $\nu_{\tau}$ appearance (bottom right) in vacuum. The legend follows that of \cref{fig:oscillation_probs_example}.}
    \label{fig:oscillation_probs_example_vacuum}
\end{figure}

\begin{figure}[htb!]
    \centering
    \includegraphics[width=0.88\textwidth]{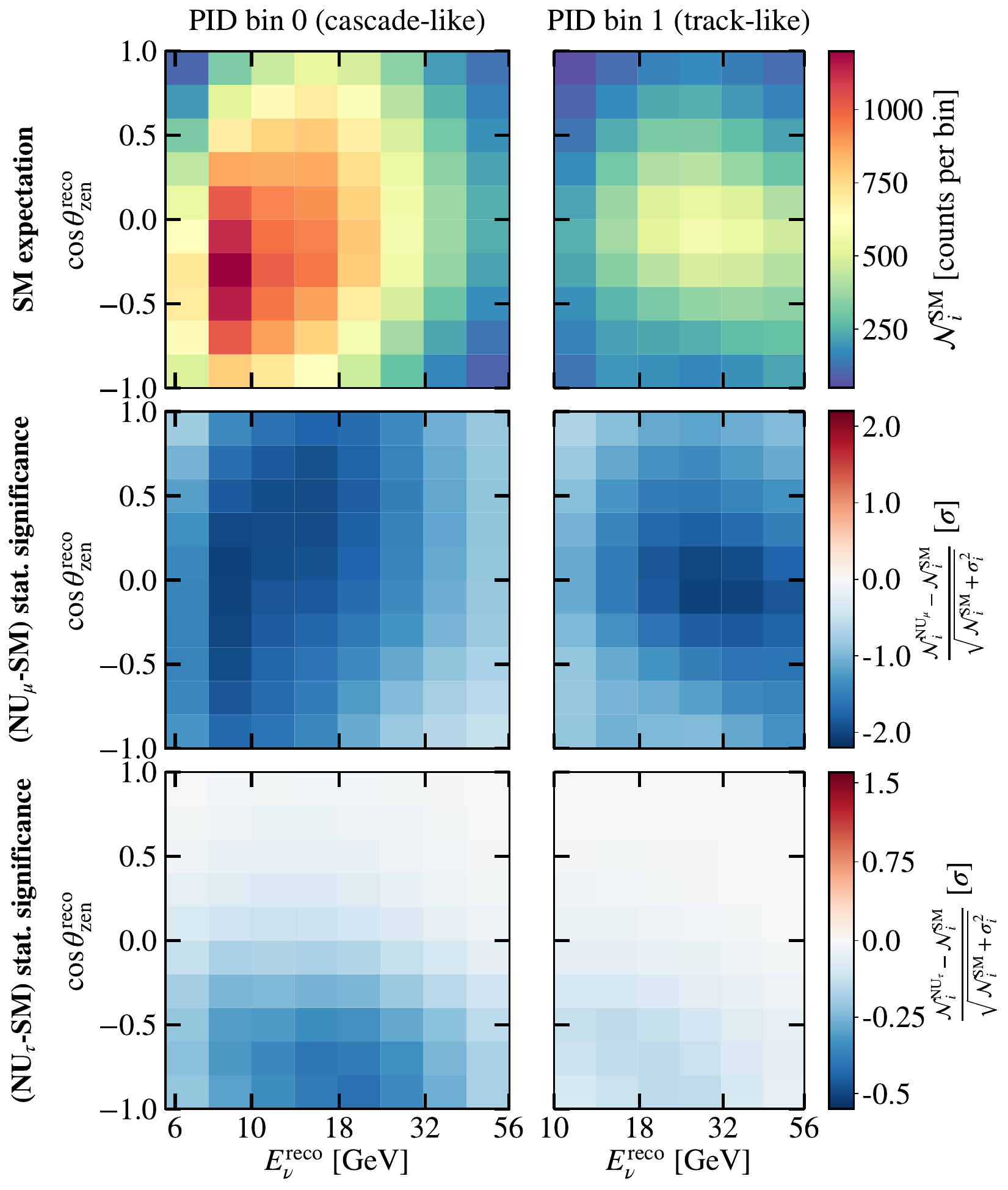}
    \caption{\textit{Top:} the unitary (Standard Model, ``SM'') expectation of the DeepCore event rates, with the same parameter setup as in \cref{fig:deepcore_offnominal_numu_templates}. \textit{Middle and bottom}: Statistical significances of the non-unitary (``NU'') expectations. All three panels assume neutrino propagation in vacuum.}
\label{fig:deepcore_offnominal_numu_templates_vacuum}
\end{figure}

\clearpage 
\section{Systematic uncertainties}

\subsection{Daya Bay covariance matrix}\label{sec:db_covariance}

In this study, we use the covariance matrix of the unfolded $\bar{\nu}_e$ spectrum from Ref.\,\cite{DayaBay:2016ssb}, which we rescale to the expected spectrum in the 3158-day analysis. As the first step, we reproduce the correlation matrix $\rho$, whose elements $\rho_{ij}$ represent the correlations between the different $\bar{\nu}_e$ energy bins $E_{\bar{\nu}_e,\,i}$ and $E_{\bar{\nu}_e,\,j}$:
\begin{equation}
    \rho_{ij} = \frac{V_{ij}^{\mathrm{DB,\,2016}}}{\sqrt{V_{ii}^{\mathrm{DB,\,2016}} V_{jj}^{\mathrm{DB,\,2016}}}},
\end{equation}
where the ``DB, 2016'' superscript refers to the covariance matrix extracted from Table 13 of \cite{DayaBay:2016ssb}. The diagonal elements of $\rho$ represent self-correlations and are equal to 1. The general relationship between the covariance matrix $V$ and the correlation matrix $\rho$ is
\begin{equation}
    V_{ij} = \rho_{ij} \sigma_i \sigma_j,
\label{eq:covariance_correlation_relationship}
\end{equation}
where $\sigma_{i(j)}$ is the absolute uncertainty on the predicted event count $\mathcal{N}_i$ in the energy bin $i (j)$. We derive this absolute uncertainty as follows:
\begin{equation}
    \sigma_i = \epsilon_i \mathcal{N}_i = \frac{\sqrt{V_{ii}^{\mathrm{DB,\,2016}}}}{\mathcal{N}_i^{\mathrm{DB,\,2016}}} \mathcal{N}_i,
\label{eq:relative_uncertainty_dayabay}
\end{equation}
where $\epsilon_i$ is the relative rate uncertainty computed from the covariance matrix element $V_{ii}^{\mathrm{DB,\,2016}}$ (Table 13 of \cite{DayaBay:2016ssb}) and the corresponding IBD rate $\mathcal{N}_i^{\mathrm{DB,\,2016}}$ (Table 12 of \cite{DayaBay:2016ssb}). Then, in the matrix form,
\begin{equation}
    V = \sum_{k} \mathrm{diag}(\epsilon \mathcal{N})_{k}  \cdot\, \rho \cdot \mathrm{diag}(\epsilon \mathcal{N})_{k},
\label{eq:covmat_matrix_form}
\end{equation}
where ``$\cdot$'' denotes matrix multiplication and $\epsilon$ multiplies $\mathcal{N}$ elementwise as per \cref{eq:relative_uncertainty_dayabay}. The index $k$ runs over the contributions of the six different reactors to the four different antineutrino detectors in EH3 and over the three data taking periods (6 AD, 7AD, and 8AD).  We assume these contributions to be uncorrelated and do not add any cross-covariance terms in \cref{eq:covmat_matrix_form}. Each of the $k$ subspectra $\mathcal{N}_k$ constitutes a fraction $\alpha_k$ of the total expected spectrum at EH3, such that $\mathcal{N}_{k} = \alpha_k \mathcal{N}^{\mathrm{exp}}_{\mathrm{EH3}}$.  Then, the covariance matrix becomes
\begin{equation}
    V = (\sum_{k} \alpha_k^2) \Big[\mathrm{diag}(\epsilon \mathcal{N}^{\mathrm{exp}}_{\mathrm{EH3}}) \cdot \rho \cdot \mathrm{diag}(\epsilon \mathcal{N}^{\mathrm{exp}}_{\mathrm{EH3}})\Big]. 
\label{eq:covmat_final_expression}
\end{equation}
We find the factors $\alpha_k$ by requiring that they are weighted according to the distance of each reactor core to each AD, the target mass and the efficiency of the ADs, as well as imposing the normalisation condition $\sum_k \alpha_k = 1$. We obtain $\sum_{k} \alpha_k^2 \simeq 0.015$, and finally use $V$ from \cref{eq:covmat_final_expression} with this prefactor to denote the systematic component $V_{\mathrm{syst}}$ of the covariance matrix in \cref{eq:chi2_daya_bay}.

We note that this is not the official approach used by the Daya Bay Collaboration, and that the above procedure reflects the steps taken in our specific study to approximate the systematic uncertainty on the expected event spectrum in the far hall of Daya Bay. We find that this treatment of the systematic uncertainties reproduces well the standard three-flavour oscillation contours from the 3158-day Daya Bay analysis, and the best-fit points in the ($\sin^2{2\theta_{13}}, \Delta m^2_{32}$) space are in excellent agreement (see \cref{fig:daya_bay_histogram}).

\subsection{Impact of the Daya Bay and KamLAND systematic uncertainties}\label{sec:db_kamland_systematics}

\begin{figure}[htb!]
    \centering
    \includegraphics[width=0.95\textwidth]{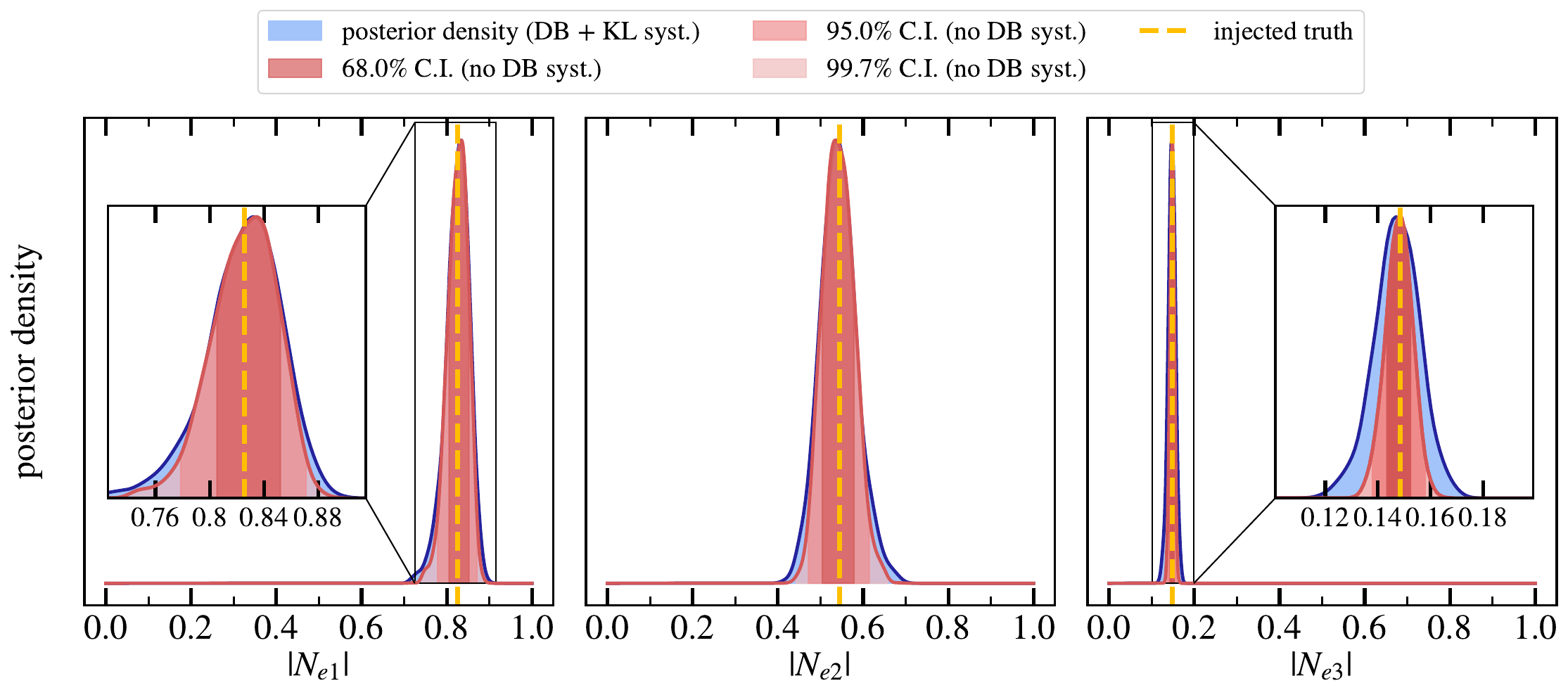}
    \caption{Projected impact of the Daya Bay flux systematic uncertainty implemented  as described in \cref{sec:db_covariance}. The blue distributions show the posteriors obtained when including both statistical and systematic components of the covariance matrix. The red distributions correspond to the case of statistical-only uncertainty. The true values of $|N_{ei}|$ (dashed yellow lines) correspond to the elements of the unitary PMNS matrix computed with NuFit 5.2 oscillation parameters \cite{Esteban:2020cvm,NuFit:2022xxx}.}
    \label{fig:dayabay_systematic_impact}
\end{figure}

\begin{figure}[htb!]
    \centering
    \includegraphics[width=0.95\textwidth]{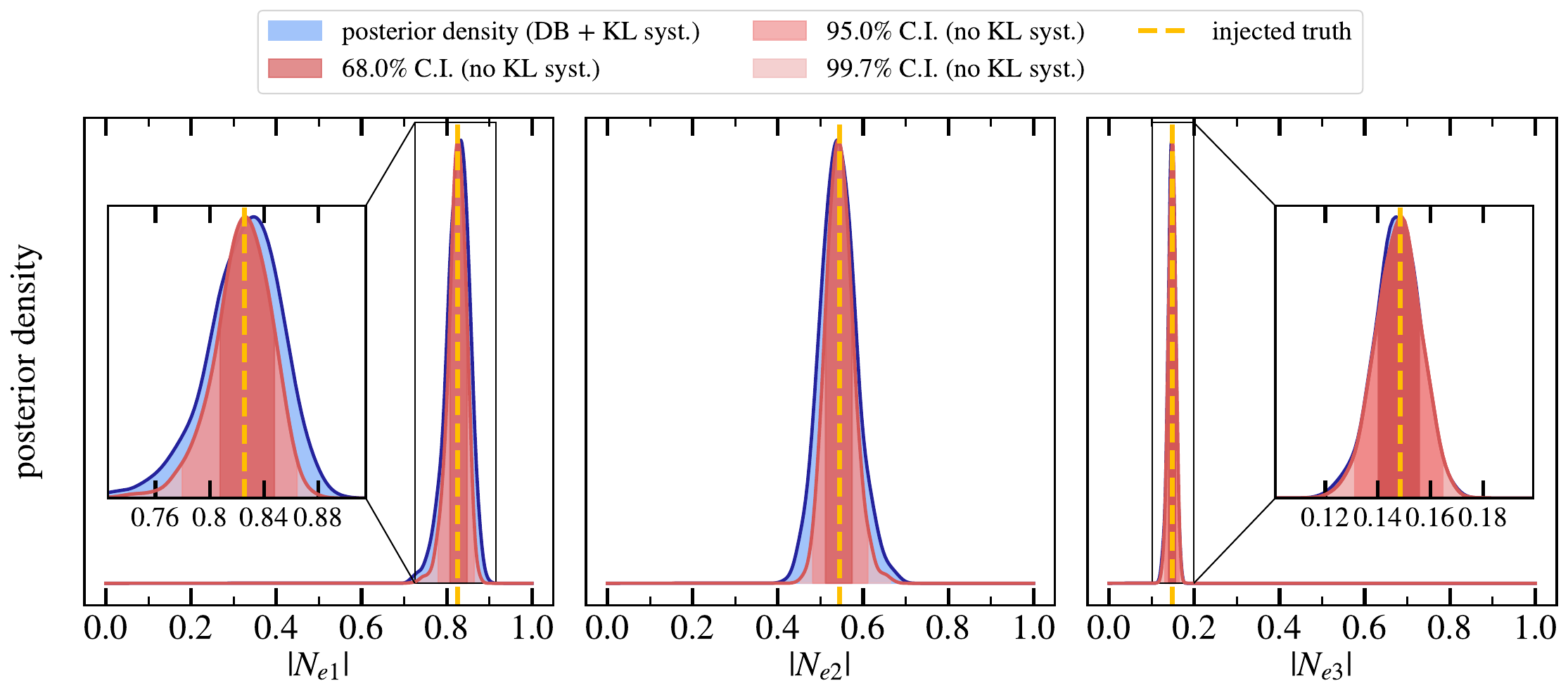}
    \caption{Same as \cref{fig:dayabay_systematic_impact}, but applied to the KamLAND flux normalisation ($\epsilon_{\mathcal{N}}$) and energy scale ($\epsilon_{E}$) systematic uncertainties.}
    \label{fig:kamland_systematic_impact}
\end{figure}

\clearpage
\subsection{IceCube-DeepCore and IceCube-Upgrade systematic parameters}\label{sec:atmo_systematics}

\begin{figure}[htb!]
    \centering
    \includegraphics[width=\textwidth]{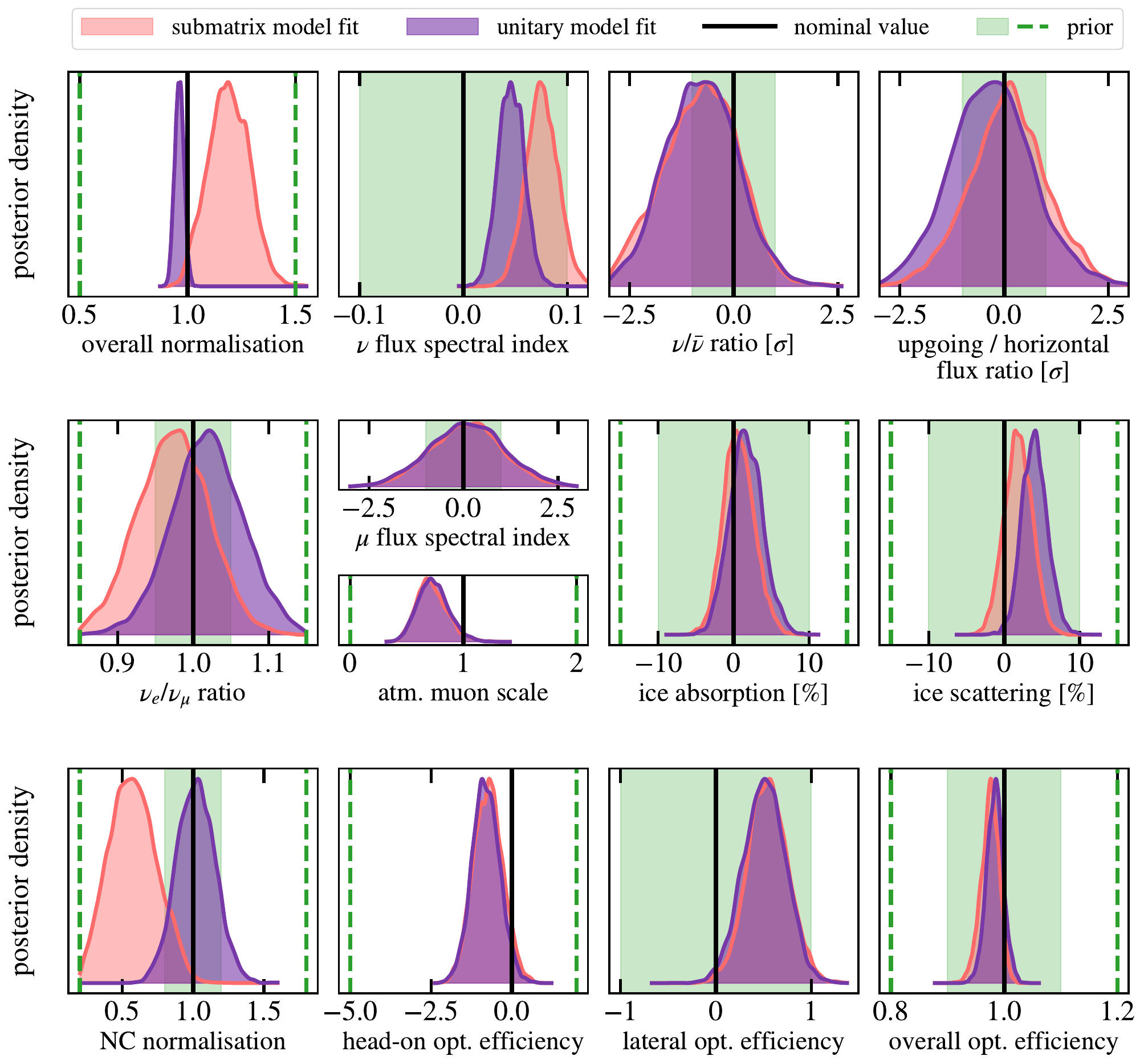}
    \caption{Posterior densities of the IceCube-DeepCore systematic parameters in the submatrix (red) and the unitary (purple) model fits to the current atmospheric + reactor neutrino data. When applicable, the rectangular shaded green area corresponds to the $1\sigma$ range of a parameter with a Gaussian prior, and the dashed green lines -- to the entire allowed range of this parameter. When only the dashed lines appear in a given panel, they represent the allowed range of a parameter with a uniform prior. The definitions and the units of the parameters follow \cite{IceCube:2019dqi}.}
\label{fig:current_sys_params_fit_results}
\end{figure}

\begin{figure}[htb!]
    \centering
    \includegraphics[width=\textwidth]{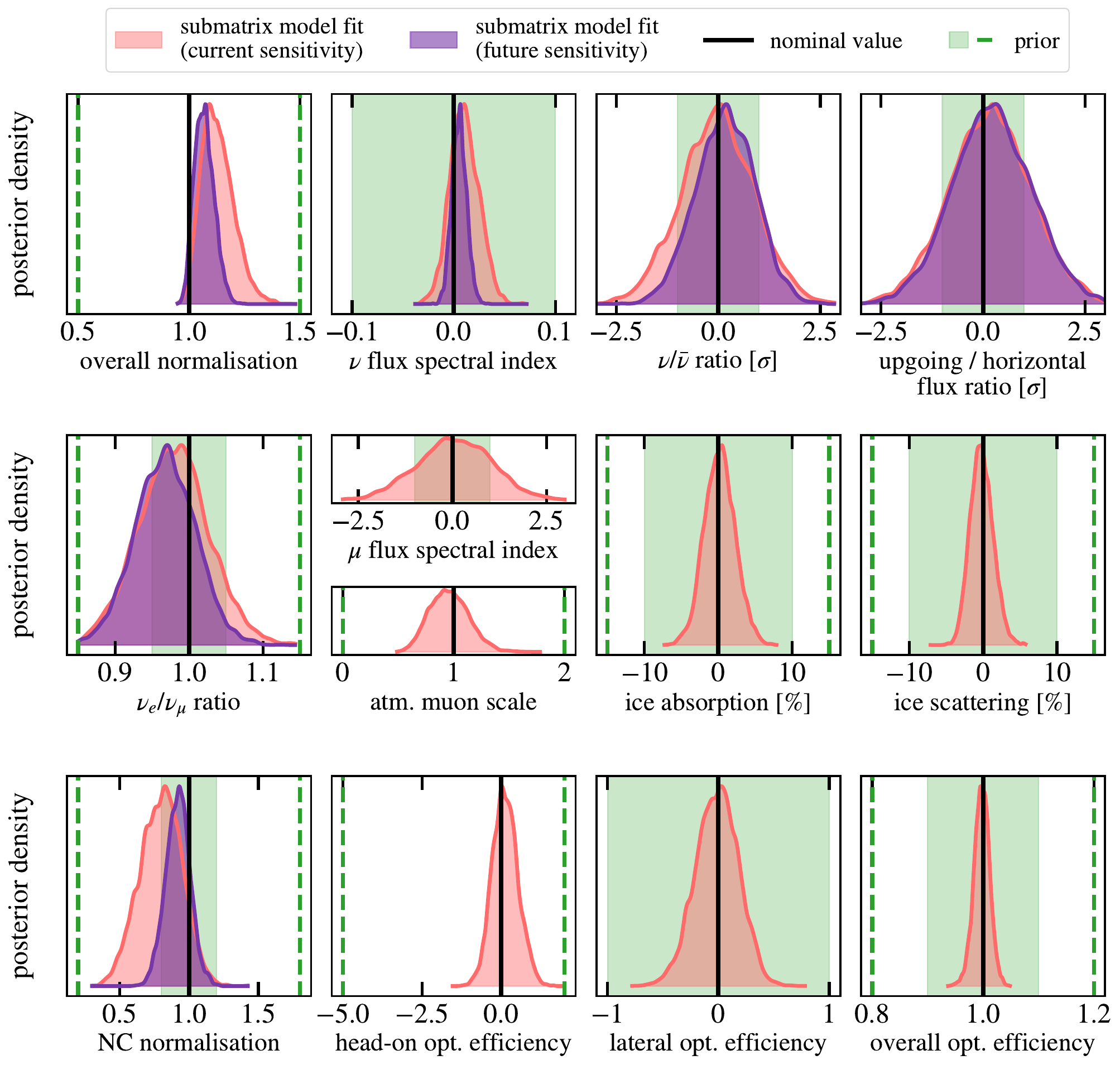}
    \caption{Posterior densities of the IceCube-DeepCore (red) and the IceCube-Upgrade (purple) systematic parameters in the submatrix model fit to the pseudodata generated assuming unitarity and NuFit 5.2 oscillation parameters \cite{Esteban:2020cvm,NuFit:2022xxx}. The priors of the parameters are shown as the rectangular shaded area ($1\sigma$ range) and the dashed lines (full range), analogously to \cref{fig:current_sys_params_fit_results}. The definitions and the units of the parameters follow \cite{IceCube:2019dqi}. If no parameter posterior is shown for the IceCube-Upgrade in a given panel, this parameter has not been included in the IceCube-Upgrade projections.}
\label{fig:current_future_sys_params_fit_results}
\end{figure}

\begin{figure}[htb!]
    \centering
    \includegraphics[width=\textwidth]{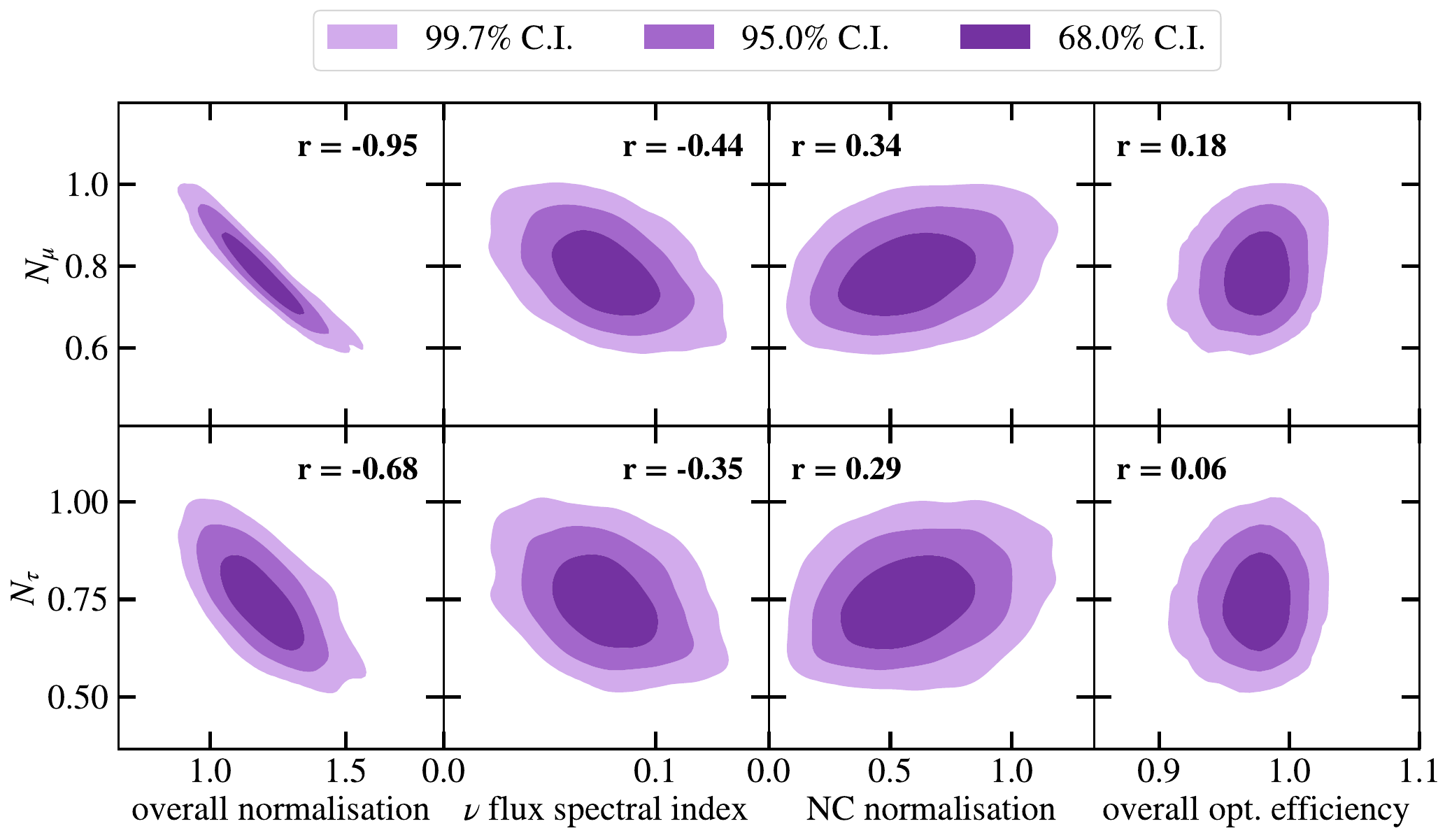}
    \caption{Correlations between the posteriors of the muon and tau row normalisations ($N_{\mu}$ and $N_{\tau}$) and a subset of the IceCube-DeepCore systematic parameters, as found in the Bayesian fit of the non-unitary mixing model (submatrix case) to the current atmospheric + reactor neutrino data. Each panel reports the Pearson correlation coefficient value ($r$) between $N_{\mu}$ or $N_{\tau}$ and a given systematic parameter, and larger absolute values of $r$ indicate stronger correlations.}
\label{fig:nmu_ntau_correlations_with_sys_params}
\end{figure}

\clearpage

\section{Future sensitivity projections for column normalisations and closures}\label{sec:future_projections_appendix}

\begin{figure}[htb!]
    \centering
    \includegraphics[width=0.75\textwidth]{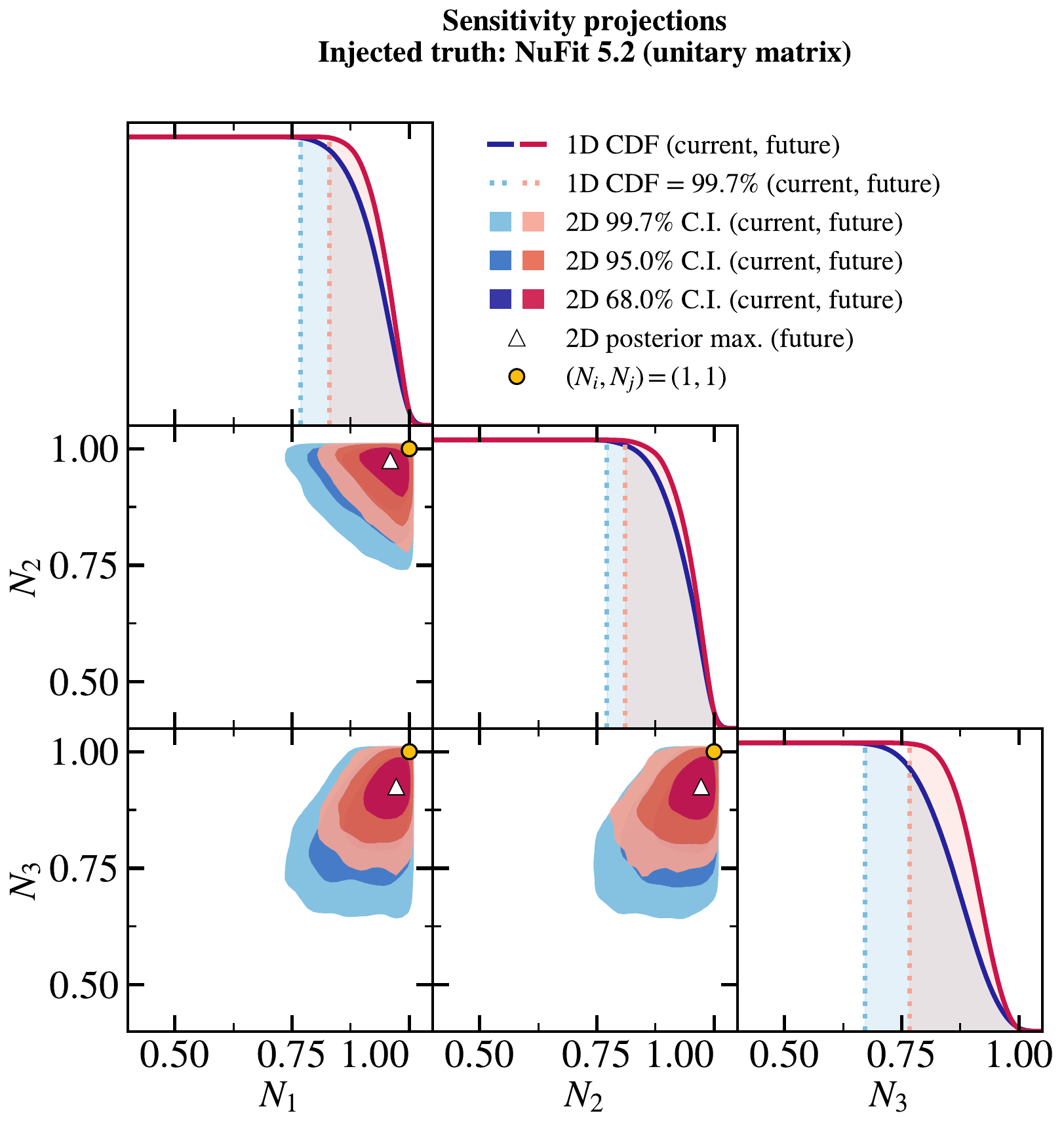}
    \caption{Same as \cref{fig:row_norms_current}, but applied to the future projections of the matrix column normalisations using the combined IceCube-Upgrade, JUNO, and Daya Bay data. The injected truth is the unitary matrix generated with NuFit 5.2 oscillation parameters \cite{Esteban:2020cvm,NuFit:2022xxx}.}
    \label{fig:col_norms_future}
\end{figure}

\begin{figure}[h!]
    \centering
    \includegraphics[width=0.75\textwidth]{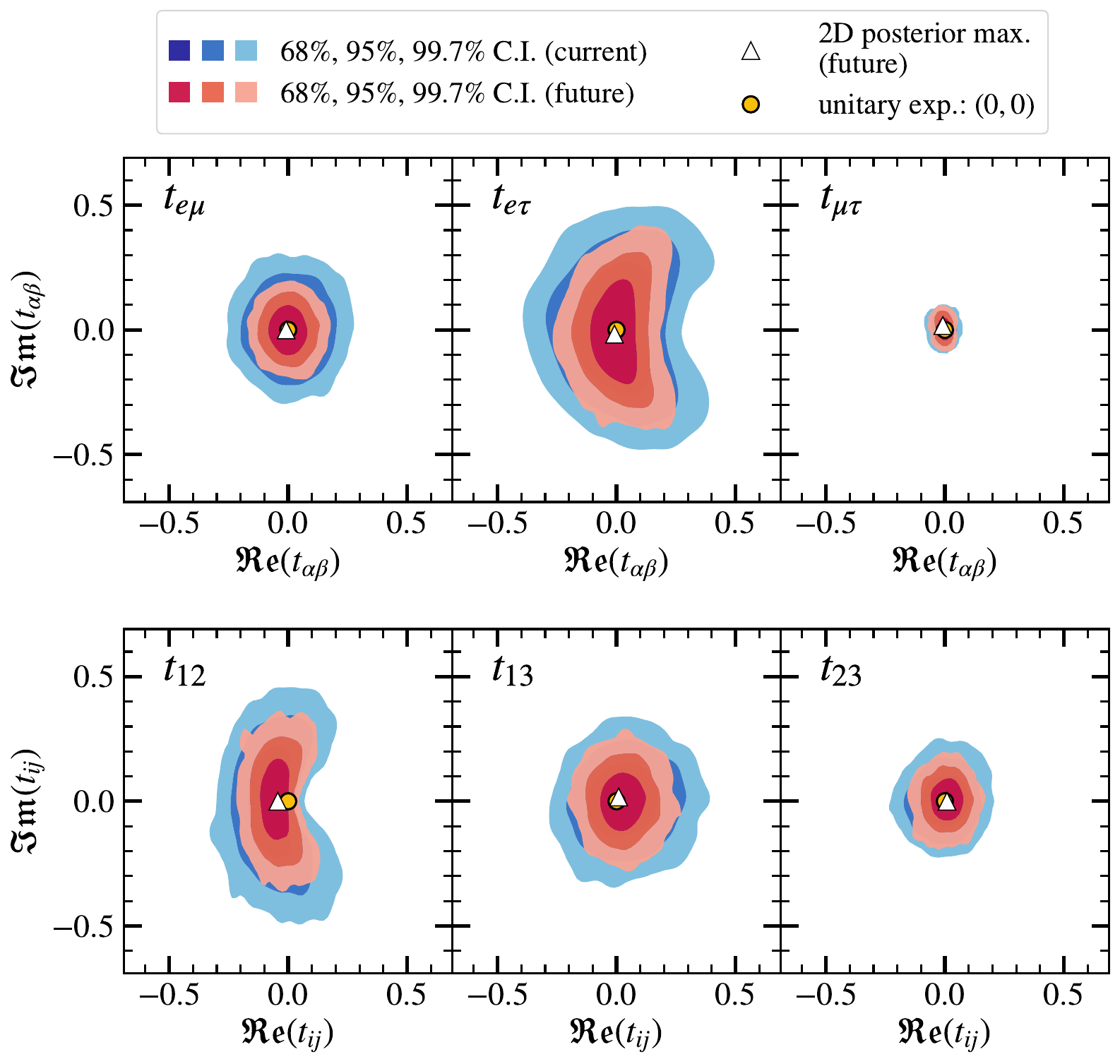}
    \caption{Same as \cref{fig:closures_current}, but applied to the future projections of the matrix row and column closures using the combined IceCube-Upgrade, JUNO, and Daya Bay data. The injected truth is the unitary matrix generated with NuFit 5.2 oscillation parameters \cite{Esteban:2020cvm,NuFit:2022xxx}. }
    \label{fig:closures_future}
\end{figure}


\bibliographystyle{JHEP}
\bibliography{main.bib}

\end{document}